\documentclass[%
 reprint,
 superscriptaddress,
 amsmath,amssymb,
 aps,
]{revtex4-2}

\usepackage{graphicx}
\usepackage{dcolumn}
\usepackage{bm}
\usepackage{hyperref}
\usepackage{times}

\newcommand*\diff{\mathop{}\!\mathrm{d}}

\begin{document}

\preprint{APS/123-QED}

\title{Arrested coarsening, oscillations, and memory from a conserved phase separating nucleator in a self-straining cytoskeletal network}

\author{Quentin Bodini-{}-Lefranc}
\affiliation{Institute of Applied Physics, TU Wien, Lehargasse 6, 1060 Vienna, Austria}
\affiliation{École polytechnique, Institut Polytechnique de Paris, Route de Saclay, 91120 Palaiseau, France}
\affiliation{contributed equally}
\author{Jakob Schindelwig}
\affiliation{Institute of Applied Physics, TU Wien, Lehargasse 6, 1060 Vienna, Austria}
\affiliation{contributed equally}
\author{Daniel Weidinger}
\author{Lucas Engleder}
\author{Sebastian F\"urthauer}
\email{fuerthauer@iap.tuwien.ac.at}
\affiliation{Institute of Applied Physics, TU Wien, Lehargasse 6, 1060 Vienna, Austria}

\date{August 3, 2025}

\begin{abstract}
 How do phase separated cellular structures set their size? To elucidate this, we study the dynamics and steady states of a phase separating nucleator that is advected by the self-straining cytoskeletal network which it nucleates. We find (i) that the interplay between transport and the tendency of nucleators to phase separate arrests coarsening; (ii) that the system undergoes damped oscillations towards a patterned steady state with a well defined length scale; (iii) that the system supports a spectrum of patterned states of different steady length scales, enabling the retention of a mechano-chemical memory of the initial conditions. Together, our findings establish a physiologically plausible compound material made of the phase separating nucleator and the self-straining network as a paradigm for mechano-chemical scale selection and memory fixation in cells. 
\end{abstract}

\maketitle

\section{Introduction}
\begin{figure*}[t]
    \centering
    \includegraphics[width=\linewidth]{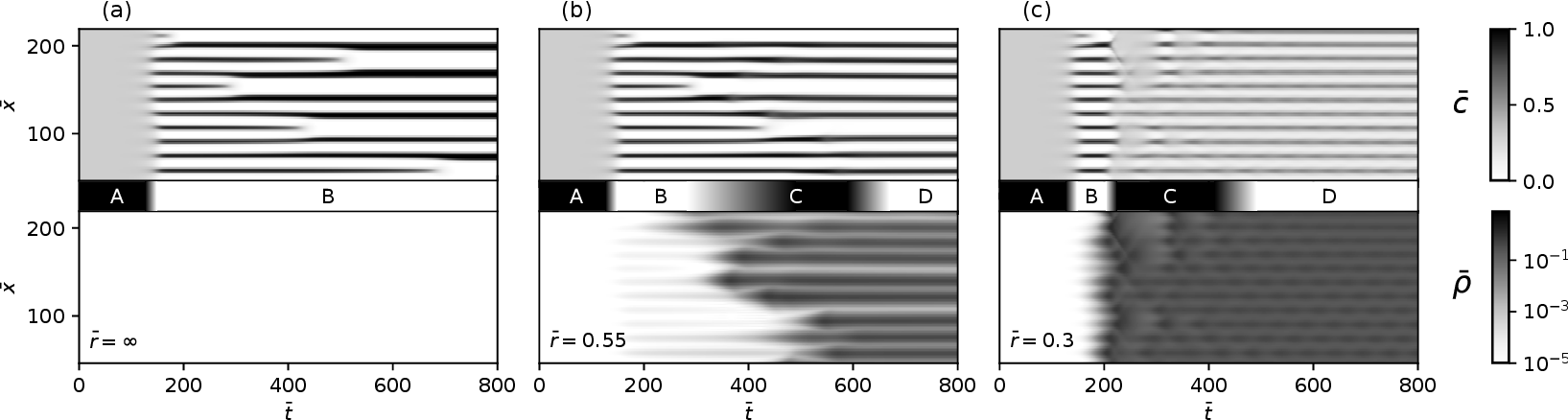}
    \caption{Kymographs of the system dynamics. Length and time scales are defined by $k$ and $k/V_{||}$, respectively, such that $\bar{x} = x/k$ and $\bar{t} = t ~ V_{||}/k$, are dimensionless length and times. The full $\bar x$-span is 400, with periodic boundary conditions. Simulation parameters are: $\bar \Lambda = \Lambda / (kV_{||}) = 1.6$; $\bar a = a / k^2 = 10/3$; $b = 1$; $\bar \kappa = \kappa / k^2= 1.75$; $\bar c_0 = c_0 k =0.3$; $\bar c_\text{c} = 0.5$; $\bar \alpha = \alpha / V_{||} = 0.9$; and $\bar \rho_\text{init} = \rho_\text{init} k = 10^{-5}$ ($\bar r = r ~ k/V_{||}$ thus defines $\bar \alpha_0$). 
    Top: $\bar c$ with a linear scale; Bottom: $\bar \rho$ with a logarithmic scale. (a) $\bar r= +\infty$: i.e. Cahn-Hilliard behaviour. (b) $\bar r = 0.55$ and (c) $\bar r = 0.3$. The letters A, B, C and D refer to phases in the behaviour of our system. A: initial $q_\text{CH}$ mode exponential growth. B: coarsening process ($q$ decreases). C: damped oscillations. D: final steady state. Note the final oscillations for $\bar r = 0.3$ in (c). Note also, that all three simulations start from the same initial random seed, for easier comparison.}
    \label{fig:simu}
\end{figure*}

Being the right size is important \cite{haldane1926being, ginzberg2015being, banerjee2025design}. Life has evolved strategies for size control on cellular scales, including phase separation \cite{boeynaems2018protein, mitrea2016phase, zwicker2025physics, weber2019physics} and reaction diffusion mechanisms \cite{turing1952chemical, kondo2010reaction, howard2011turing, gessele2020geometric}. 
Equilibrium phase separation and mass-conserving reaction-diffusion processes (which can be mapped to the Cahn Hilliard (CH) equation \cite{bergmann2018active, robinson2025universal}) generically end in full coarsening \cite{cahn1958free, cahn1965phase, novick1985nonlinear, bray2003coarsening, ostwald1903lehrbuch, brauns2021wavelength, weyer2023coarsening, gonnella2015motility}, and thus generate a single structure whose size is set by the abundance of a limiting component.
Yet, in cells many phase separated structures can coexist, hinting at physics beyond equilibrium. Indeed, coupling CH to a detailed-balance-breaking chemistry has emerged as a common motif for achieving long-time length scale selection \cite{zwicker2015suppression, zwicker2022intertwined}. In this paper we are concerned with a second physiologically plausible strategy for injecting energy, which is via active mechanical processes \cite{garcke2003cahn, padhan2024novel, frohoff2021suppression, frohoff2023non}.
We show that suppression of coarsening [Fig.~\ref{fig:simu}] and length scale selection [Fig.~\ref{fig:q_selec}] naturally emerge from coupling CH to active transport by a self-straining network.
Further, we find that in the same system patterned states of different steady length scales coexist [Fig.~\ref{fig:stab}], allowing the compound material of CH nucleator and self-straining cytoskeletal network to act as a cellular memory. 

In cells, active mechanics and nucleation conspire to size organelles \cite{mao2014behaviour, chen2023regulation}. However, the physics governing this selection have remained largely opaque. 
Cytoskeletal networks are the mechanical bones and muscles of a cell. They set the shape of cells and allow them to deform, divide and rearrange \cite{alberts2008molecular}. These networks are made from structurally polar biopolymer filaments (actin, microtubules) connected by motorized crosslinks (kinesin and dynein, myosin). Filaments and (motorized) crosslinks typically turn over rapidly compared to large-scale rearrangement \cite{needleman2010fast, fritzsche2013analysis}, thus arguing that length scale control of cytoskeletal structures hinges on an interplay between active mechanics and nucleation. The inner workings of this interplay are yet to be understood. To address this problem we formulate a minimal model for a highly crosslinked cytoskeletal network of mass density $\rho$ and polarity $P$, nucleated by a phase-separating species of density $c$. For simplicity we will discuss the one-dimensional case.

\section{Results}
\subsection{Emergence of an active self-straining network arrests Cahn-Hilliard coarsening}
\label{sec:eq}

The equations of the cytoskeletal network read
\begin{subequations}
\label{eq:cyto}
\begin{eqnarray}
\partial_t \rho & = & (\alpha c - r) \rho + V_{||}\partial_x (\rho P) + \alpha_0, \label{eq:rho}\\
\partial_t (\rho P) & = & -r \rho P + V_{||}\partial_x \rho,
\end{eqnarray}
\end{subequations}
where $V_{||}$ is the filaments sliding speed, $r$ is the turnover rate, and $\alpha$ the branching nucleation rate \cite{petry2013branching, kaye2018measuring}. Finally, $\alpha_0\simeq 0$ is a small spontaneous production term, that we choose negligibly small in what follows. The derivation of Eqs. (\ref{eq:cyto}) closely follows \cite{furthauer2019self, furthauer2021design, zampetaki2025}; see Appendix \ref{app:A}. The nucleator density $c$ dynamics obeys a CH equation \cite{cahn1958free} coupled to the cytoskeletal network through advection:
\begin{equation}
\label{eq:c}
    \partial_t c = \partial_x\big(\Lambda \partial_x(\mu(c))\big) + k V_{||} \partial_x (\rho P c),
\end{equation}
where $\mu(c) = a(c-c_\mathrm{c})^3 - b(c-c_\mathrm{c}) - \kappa \partial_x^2 c$ is a chemical potential with critical concentration $c_\text{c}$ \cite{ginzburg1950theory}, $\Lambda$ is a constant mobility, and $k$ sets the advection strength.

We solve Eqs. (\ref{eq:cyto}, \ref{eq:c}) numerically using the spectral package Dedalus \cite{burns2020dedalus}. Fig. \ref{fig:simu} shows simulation results for three cases which differ in the actin turnover rate $r$. The Code, and the input fies are available in the repository \cite{CodeAcc}. In contrast to pure CH dynamics [Fig. \ref{fig:simu}(a)]---where an initial length scale is selected (phase A) before coarsening through Ostwald ripening occurs (phase B)---coarsening is suppressed at long times when coupled to the active network [Fig. \ref{fig:simu}(b, c)]. Coarsening suppression coincides with the growth of this active network, and damped oscillations in $\rho$ and $c$ (phase C) are observed as the system settles into a periodic structure of defined length scale (phase D). We first seek to understand how this length scale is selected.

\begin{figure}[t]
    \centering
    \includegraphics[width=\linewidth]{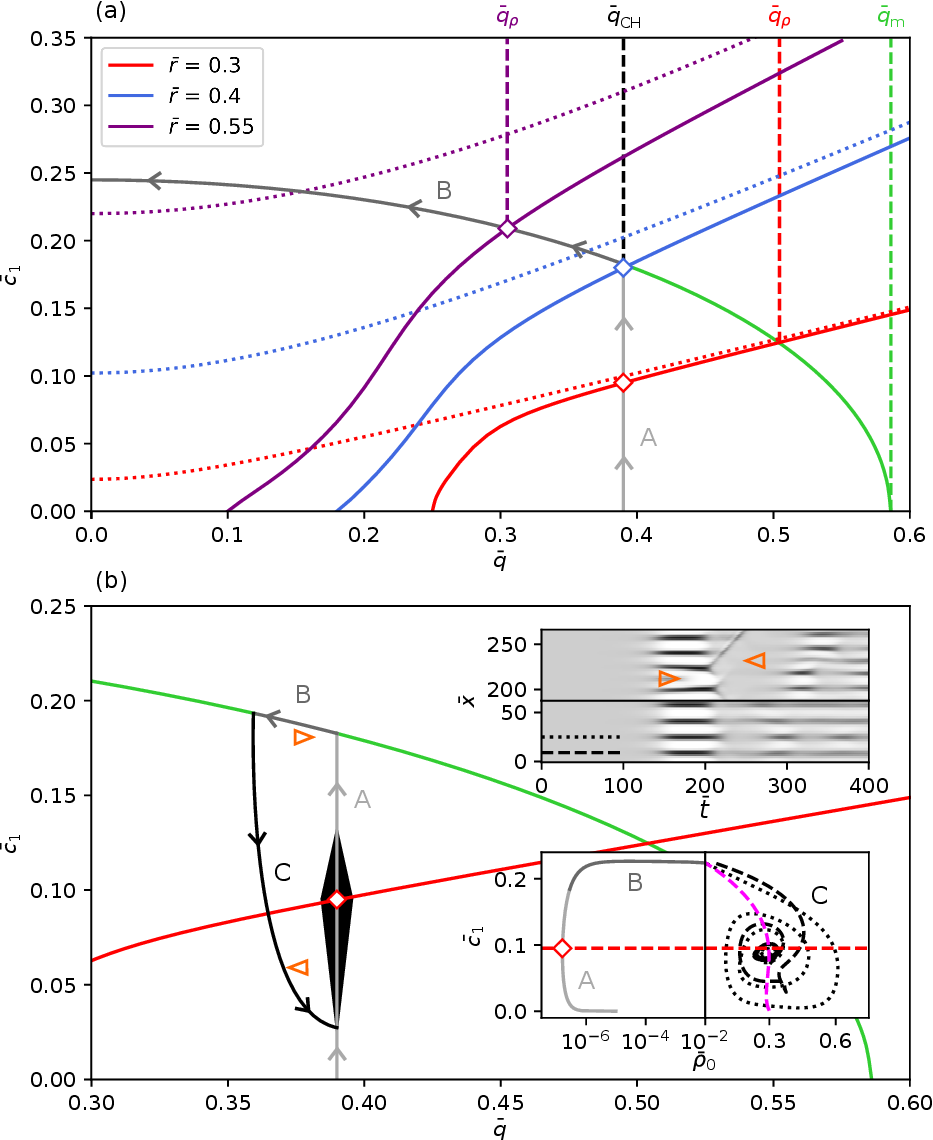}
    \caption{(a) Green: $c_{1,\text{CH}}$ obtained from Eq. (\ref{eq:amplitude}). Red, Blue, Purple: $c_{1,\text{th}}$ obtained from Eq. (\ref{eq:c_th}) in dotted lines, and correction with further harmonics using Eq. (\ref{eq:tilde}) in solid lines. The gray trajectory is pure CH [Fig. \ref{fig:simu}(a)], and diamonds indicate when the real trajectory diverges from it.
    (b) Real trajectory for $\bar r=0.3$. The orange triangles refer to isolated coarsening and reverse coarsening events. Note the damped oscillations around the diamond (better seen in the lower inset). In the upper inset, extracts from Fig. \ref{fig:simu}(c). In the lower inset, phase diagram in the $(\rho_0, c_1)$ plane. Note that $\bar \alpha_0=0$ and $\bar \rho_\text{init}=10^{-5}$ to better see the onset of $\rho$. Red: $c_{1,\text{th}}$, Magenta: $\rho_{0,\text{th}}$. Logarithmic scale for $\rho_0 < 10^{-2}$, linear scale above. Two trajectories are drawn, from two different crests in the upper inset (dotted or dashed line).}
    \label{fig:q_selec}
\end{figure}

\subsection{Length scales are selected by Cahn-Hilliard and actin nucleation kinetics} 
\label{sec:scales}
The simulation in Fig. \ref{fig:simu} starts from the initial condition $\rho_\text{init} = \alpha_0 / (r - \alpha c_0)\simeq 0$, $P=0$, and $c = c_0$, with a small white noise perturbation on $c$. Importantly $c_0 < r / \alpha$, thus advection can initially be ignored in Eq. (\ref{eq:c}). A typical CH behavior for $c$ follows: (phase A) a space scale with wavenumber $q_\text{CH} = \sqrt{(b-2a(c_0-c_\text{c}))/2\kappa}$ is selected \cite{cahn1965phase, novick1985nonlinear}, and exponentially grows until an unstable spatially periodic steady state $c_\text{CH}(q_\text{CH})$ is reached; (phase B) Ostwald ripening ensues and the system progresses through periodic steady states $c_\text{CH}(q)$ with decreasing wavenumbers $q$. In pure CH this coarsening process would continue until complete phase separation. In our simulations however, the behavior of the system departs from pure CH after some initial coarsening and the active network $\rho$ starts to grow.
To understand when this happens, we will use the steady state structure of CH. Thus, before discussing the onset of phase C, we need to review the structure of CH stationary states.

Following the methods of \cite{langer1971theory,novick1984nonlinear, argentina2005coarsening} the profile $c_\text{CH}(q)$ can be derived for any given $c_0$ and $q \in [0,q_\text{m}]$. Here $q_\text{m}$ is the largest unstable wavenumber for the CH instability. This yields the approximation
\begin{subequations}
\label{eq:steady}
\begin{eqnarray}
    c_\text{CH} &=& A + \delta - \frac{4A\delta+2\delta^2}{2A +\delta + (2A-\delta) ~ \mathrm{sn}^2\left(K x, m\right)}\\
    &=& c_0 + 2\sum\limits_{n\geq1} c_{n,\text{CH}} \cos(nqx), \label{eq:fourier}
\end{eqnarray}
\end{subequations}
where $A$, $\delta$, $K$ and $m$ are functions of $q$, $c_0$, $b/a$ and $\kappa/a$, and $\mathrm{sn}$ is a Jacobi elliptic function; see Appendix \ref{app:B.1}. The Fourier coefficients $c_{n,\text{CH}}$, in Eq.~{(\ref{eq:fourier})} can be computed by writing Eq. (\ref{eq:c}) as a hierarchy of equations for all $c_n$. Using that $|c_{n,\text{CH}}|$ scale as $\exp(-n)$, we truncate the equations of this hierarchy to express the dynamics of $c_n$ (around the stationary state) as functions of the lower harmonics $c_{i \leq n}$ of $c$ only; see Appendix \ref{app:B.2}. The first of these harmonic amplitude equations encodes for the dynamics of $c_1$ and only depends on $q$, $c_0$ and $c_1$ itself. It reads
\begin{equation}
\label{eq:amplitude}
    \partial_t c_1 = \Lambda q^2 \left( (b-3a(c_0-c_\text{c})^2-\kappa q^2) c_1 - 3 a c_1^3 \right).
\end{equation}
Setting Eq. (\ref{eq:amplitude}) to zero yields an approximation of $c_{1,\text{CH}}$ given $q$. 
The other equations of the hierarchy (set to zero) slave the dynamics of $c_{n \geq 2}$ to $c_1, c_0$ and $q$, effectively providing a map
\begin{equation}
    (c_{1},q) \mapsto  c_{n\geq2}. \label{eq:tilde}
\end{equation}
Thus, for any given wavenumber $q$ and amplitude $c_1$, Eq.{(\ref{eq:tilde})} provides a $c$ profile that respects the structure of CH stationary states. For the special case of $c_1=c_{1,\text{CH}}$ we recover $c_n = c_{n,\text{CH}}$ and thus the profile $c_\text{CH}$. 
Beyond pure CH, Eq.~\ref{eq:tilde} sets us up to determine the conditions for growth of $\rho$, which eventually brings phase B to an end:
with Eq.~(\ref{eq:tilde}) the higher harmonics $c_{n \geq 2}$ are functions of $c_1$ and $q$ and thus, we can now determine a $q$-dependent threshold for $c_1$ beyond which the active network starts to appear.
 
For an arbitrary profile of $c$, Eqs. (\ref{eq:cyto}) can be written in the Fourier space as
\begin{subequations}
\label{eq:hierarchy}
\begin{eqnarray}
    \partial_t \rho_n & = & \alpha \sum_m \left( c_{n-m} \rho_m \right) - r \rho_n + \mathrm{i}  n q V_{||} \Phi_n,\\
    \partial_t \Phi_n & = & -r \Phi_n +  \mathrm{i} n q V_{||} \rho_n,
\end{eqnarray}
\end{subequations}
where we introduced $\Phi = \rho P$, for notational convenience. 
The subscripts $n$ also denote Fourier coefficients for $\rho$ and $\Phi$. Let us first consider the profile of $c$ during phase A: the $q_\text{CH}$ mode grows exponentially, hence $c$ is nearly sinusoidal and the $c_{n\geq2}$ are negligible. Eqs. (\ref{eq:hierarchy}) thus yields a threshold $c_{1,\text{th}}(q_\text{CH})$ beyond which $\rho$ grows:
\begin{equation}
\label{eq:c_th}
c_{1,\text{th}}(q_\text{CH}) = \frac1{2\alpha}\sqrt{2(r-\alpha c_0)\left(r-\alpha c_0 + \frac{q_\text{CH}^2V_{||}^2}{r}\right)}.
\end{equation}

Therefore, if $c_{1,\text{th}}(q_\text{CH}) < c_{1,\text{CH}}(q_\text{CH})$, the growth of $\rho$ starts during phase A. This is the case for our simulations with $\bar r = 0.3$ [Fig. \ref{fig:simu}(c) and \ref{fig:q_selec}(a)]. In this case, phase B is a short transient phase during which $\rho$ exponentially grows from small $\rho_\text{init}$. 

If $c_{1,\text{th}}(q_\text{CH}) > c_{1,\text{CH}}(q_\text{CH})$ the situation becomes more complex, since now the threshold of Eq. (\ref{eq:c_th}) is not crossed at $q_\text{CH}$. The Ostwald ripening during phase B is then longer and of greater importance: the growth of $\rho$ sets in only after sufficient coarsening. This is the case for the simulations with $\bar r=0.55$ [Fig. \ref{fig:simu}(b) and \ref{fig:q_selec}(a)]. The criterion obtained in Eq. (\ref{eq:c_th}) is no longer strictly valid, since $c$ has departed from its sinusoidal shape, but instead follows the profiles given by Eqs. (\ref{eq:steady}). Using Eq. (\ref{eq:tilde}) and seeking a zero eigenvalue for Eqs. (\ref{eq:hierarchy}), we obtain a threshold $c_{1,\text{th}}(q)$ [Fig. \ref{fig:q_selec}(a)]; see Appendix \ref{app:C}. 

The physical picture that emerges form this analysis is the following: when $c >  r / \alpha$ in parts of the domain, $\rho$ may start to grow locally [see Fig. \ref{fig:simu}(b) for $\bar t \approx 40$]. Two effects are competing: (i) the first term of Eq. (\ref{eq:rho}) is now positive, which favors the growth of $\rho$; (ii) when growing, $\rho$ inherits the gradients of $c$, hence the second term of Eq. (\ref{eq:rho}) becomes important and leads to a spreading out of $\rho$. This generically lowers peaks of $\rho$ and inhibits $\rho$-growth near sharp interfaces. Thus, whether $\rho$ grows or not is set by the amplitude of $c$ and by its gradients, which explains the $q$-dependence of $c_{1,\text{th}}$. 
Given that $c_{1,\text{th}}$ and $c_{1,\text{CH}}$ are $q$-dependent, two distinct situations can occur: either the threshold is crossed during the initial CH phase separation (phase A) (if $c_{1,\text{th}}(q_\text{CH}) < c_{1,\text{CH}}(q_\text{CH})$); or the threshold will only be crossed in phase B, after some Ostwald ripening has happened (if $c_{1,\text{th}}(q_\text{CH}) > c_{1,\text{CH}}(q_\text{CH})$). The wavelength of the patterns at which the system departs from pure CH behavior are therefore different in these two regimes: in the former the selected wavenumber is $q_\text{CH}$, while in the latter it is $q_\rho$, defined as the maximal wavenumber for which $c_{1,\text{th}}(q) < c_{1,\text{CH}}(q)$. 
The different scenarios are displayed in Fig. \ref{fig:q_selec}(a). 

\subsection{Damped oscillations appear generically before the system reaches a steady patterned state}
We next seek to understand how the emergence of an active network feeds back onto the dynamics of $c$ and explain the new patterned steady states observed in Fig. \ref{fig:simu}. We start this analysis by noting that  $\rho$ and $\Phi$ grow with rates $\alpha c - r$ and $-r$, respectively; see Eqs. (\ref{eq:cyto}). Thus around the threshold $\alpha c -r \simeq 0$ the dynamics of $\Phi$ is fast compared to that of $\rho$ and an adiabatic approximation $\Phi \simeq V_{||} / r ~ \partial_x\rho$ can be made. In our simulations this  approximation remains valid throughout; see Appendix \ref{app:D.1}. Further, in our analysis we will assume that to leading order only the dynamics of $c_1$ are modified by the presence of the active network, i.e. that we can still write higher harmonics as functions of $c_1$ and $q$, using Eq. {(\ref{eq:tilde})}; this assumption is justified in Appendix \ref{app:D.2}. 
With these assumptions, the condition for a steady $c$ is simply $\partial_t c_1 =0$. Using Eq. (\ref{eq:c}) and going to Fourier space we can rewrite this condition as follows   
\begin{equation}
\label{eq:rho_th}
 \sum_{n \geq 1} \rho_{n} (c_{n-1} + c_{n+1}) = \frac{-r\Lambda}{kV_{||}^2q^2} \Big[\partial_x^2(\mu(c))\Big]_1(c_0,c_1,\dots),
\end{equation}
where the subscript in $[\cdot]_1$ denotes the first Fourier coefficient. Note that by using Eq.~\ref{eq:tilde} once more, this can be made into a condition on $c_1$ and the harmonics of $\rho$.

To further reduce the dimensionality of conditions Eq.~{(\ref{eq:rho_th})}, we use the structure of the profiles of $\rho$ near the threshold $c_{1, \text{th}}$. We numerically observe that near threshold, and to good approximation throughout our simulations, the $\rho$ dynamics mainly occurs along the eigenvector $\vec\rho_{\sigma=0}$ of the largest eigenvalue $\sigma=0$; see Appendix \ref{app:C}. Moreover, $\vec\rho_{\sigma=0}$ is almost parallel to $\rho_0$. This provides a center manifold reduction.
Therefore, $\rho_0$ and $q$ are sufficient to describe all the dynamically accessible $\rho$ profiles, and $\rho_{n\geq1}$ can be expressed in terms of $\rho_0$ and $q$. 
Effectively, this allows us to calculate a threshold value $\rho_{0,\text{th}}$ from Eq.~{(\ref{eq:rho_th})}, beyond which $c_1$ should start to decay. 

In the insert of Fig.\ref{fig:q_selec}(b) dashed lines indicate $c_{1,\text{th}}$ and $\rho_{0,\text{th}}$ in red and pink, respectively. Their crossing point identifies the steady state of the system. Comparing to simulations (dotted and dashed black lines) yields good agreement between analytics and numerics. Thus we have now gained an analytical understanding of the steady states of the system. We next ask whether the approach towards the steady state is also covered by our descriptions

Numerically we observe that the final state is reached after damped oscillations. These oscillations are the most  dramatic in cases where the growth of $\rho$ starts during phase A.
In this scenario, exponential growth of $c_1$ and some initial coarsening events can happen before $\rho_0$ reaches $\rho_{0,\text{th}}$ [Fig. \ref{fig:q_selec}(b)]. This leads to a short transient phase B and a large overshooting of $c_{1,\text{th}}$. Damped oscillations of large amplitude follow. The inflection points of these damped oscillations are well described by $c_{1,\text{th}}$ and $\rho_{0, \text{th}}$ derived earlier [Fig \ref{fig:q_selec}(b), lower inset]. However spatial inhomogeneities of the system now become important.
Isolated coarsening events observed during the initial growth of $\rho$ break spatial symmetries. Often---but not always---they are compensated by reverse coarsening events [see orange triangles in Fig. \ref{fig:q_selec}(b)]. Note that during these peak splitting events, $c$ material travels. The final state is therefore characterized by a dominant wavenumber which may be slightly lower than $q=q_\text{CH}$.

Furthermore, the amplitude of the damped oscillations also varies due to traveling $\rho$ material. Noise in the initial condition can lead to variations in the height of $c$ peaks after CH phase separation, and deviations from the perfect periodicity that we assumed in our analytics. As a consequence regions of more or less $c$ will produce exponentially more or less $\rho$ material, and act as sources and sinks, respectively.
This accordingly precipitates or delays the decay of $\rho$ (thereby deviating from $c_{1,\text{th}}$), and thus influences the amplitude of the subsequent damped oscillations [Fig. \ref{fig:q_selec}(b), dotted and dashed lines in insets].
It also explains the non-simultaneous transition between phase B and C observed in Fig. \ref{fig:simu}(c).

\begin{figure}[t]
    \centering
    \includegraphics[width=\linewidth]{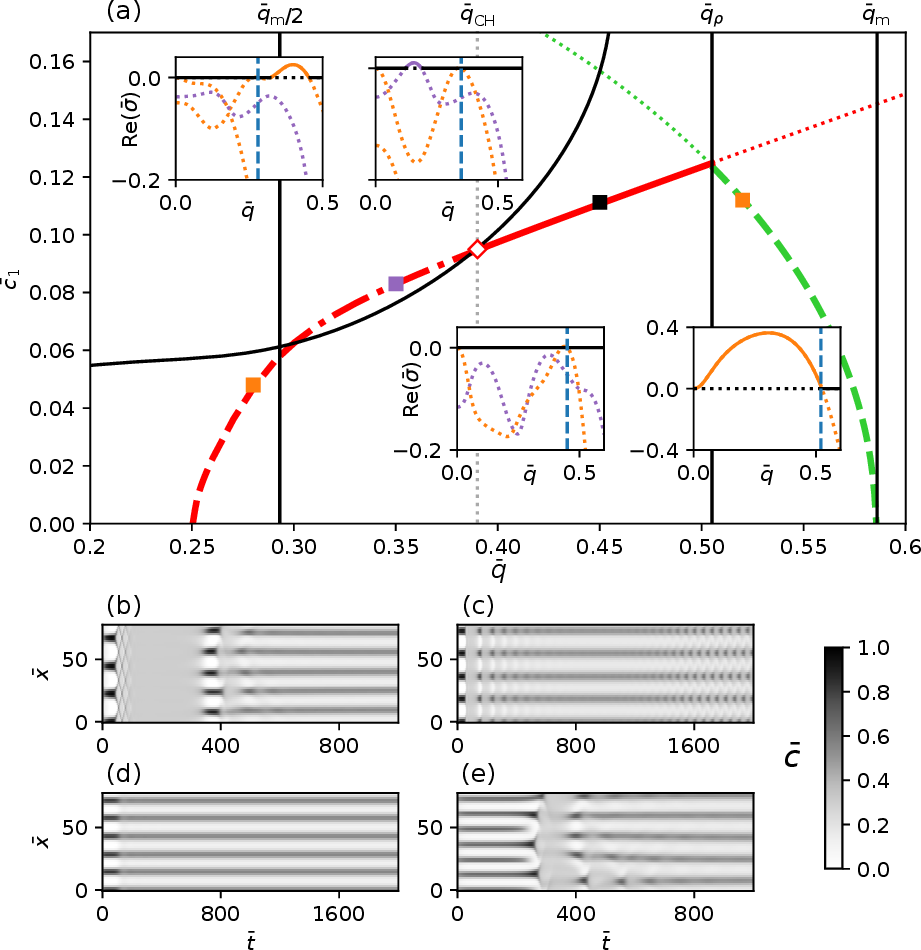}
    \caption{(a) Steady states plotted in the $(q,c_1)$ plane. Green: CH states; Red: active states with non-trivial $\rho$. Dotted: non-steady states; Dashed: stationary instability (large-scale/coarsening or short-scale/reverse-coarsening); Dot-dashed: oscillatory instability (of twice the wavelength); Solid: stable steady-state. In the insets are plotted the real part of the leading eigenvalues of the linear stability matrix, around each of the square-labeled steady states. Orange: real eigenvalue; Purple: eigenvalue with non-zero imaginary part. Blue: wavenumber $q$ of the perturbed steady state.
    (b-e) Simulated kymographs of $\bar c$ with parameter values from Fig. \ref{fig:simu} and $\bar r=0.3$ (except $\bar\Lambda = 1.25$ in (b) and $\bar\Lambda = 1$ in (e)), with sinusoidal profile of wavenumber $q$ (and small noise) for $c$, $\rho_\text{init}=\alpha_0 / (r - \alpha c_0)$, and $P=0$ as initial conditions. These wavenumbers $q$ correspond to the squares in (a). The trajectories are attracted by the associated steady state (square), before experiencing its (in)stability.
    }
    \label{fig:stab}
\end{figure}

\subsection{Steady patterns of different wavelength coexist}
In the previous paragraphs, we derived steady states with non-trivial $\rho$ and $P$ profiles for $q<q_\rho$, and discussed the state which the system reaches by following its native dynamics from a ‘naive' initial condition. However, steady states beyond this one exist. We next ask which of them are stable.

For this, we truncate Eqs.(\ref{eq:hierarchy}) and the Fourier version of Eq. (\ref{eq:c}) at $n=1$. We consider sinusoidal steady states characterized by $q$, $c_0$, $c_1$, $\rho_0$, $\rho_1$ and $\Phi_1$. Given the nonlinearities in our equations, we consider perturbation of angular wavenumbers $p$, $q+p$ and $q-p$. After linearizing around the steady state, this yields a $9\times9$ linear system; see Appendix \ref{app:E}. In Fig. \ref{fig:stab}, we show the results of this stability analysis for the different regimes.

Let us consider the different cases obtained for $\bar r=0.3$ [Fig. \ref{fig:stab}(a)]. When (i) $q_\text{m} < q$, there is no non-trivial steady state. For (ii) $q_\rho < q < q_\text{m}$, the growth of $\rho$ is not activated, and the CH state is still steady. It features a coarsening instability with a fastest growing wavenumber smaller than $q_\text{CH}$ [Fig. \ref{fig:stab}(e)].
Note that for $q>q_\rho$, there is no steady state with non-trivial $\rho$ [see magenta line in Fig. \ref{fig:q_selec}(b)]. When (iii) $q_\text{m}/2 < q < q_\rho$, the CH state is no longer steady, and the new steady state with non-trivial $\rho$ may either be stable or experience an oscillatory instability.  This oscillatory instability happens for large enough $c_1$ [see the curved black line in Fig. \ref{fig:stab}(a)]. Note that for $q_\text{CH}$, the oscillatory instability is slightly on, hence the central circle in the lower inset of Fig. \ref{fig:q_selec}(b), and the final oscillations in Fig. \ref{fig:simu}(c).
The spatial period of these final oscillations is twice that of the steady state. This hints at dynamics where nearby droplets periodically exchange material. Most importantly, for $q_\text{m}/2 < q < q_\rho$, there is neither coarsening nor reverse coarsening [Fig. \ref{fig:stab}(c, d)]. Finally, for (iv) $q<q_\text{m}/2$, one can see a reverse coarsening instability. This leads to mesa splitting events [Fig. \ref{fig:stab}(b)].
Note that in the region  $\bar q\lesssim 0.25$, where the red line goes to zero, some of the approximations made in our calculations break down, and thus our analytical understanding is limited. In simulations of this regime, we observe reverse coarsening instabilities for arbitrarily small $q$, with final states always featuring $q_\text{m}/2 < q < q_\rho$.

Importantly, steady states within the interval $[q_\text{m}/2, q_\rho]$ preserve their wavelength.
In this regime, the system supports a broad spectrum of patterned states with different wavelengths. Thus information about the systems past can be encoded in its final state. The system displays memory.

\section{Discussion} 
Here we have shown that the coarsening process of CH can be stopped when coupled to a self-straining cytoskeletal network. This coupling (i) dynamically selects a length scale for the a patterned final state; (ii) is accompanied by damped oscillations towards the steady state; (iii) supports a broad spectrum of stable steady states, such that memory about the systems history can be stored.

Importantly, Eqs. (\ref{eq:cyto}, \ref{eq:c}) are biologically plausible. Sliding of cytoskeletal filaments exists in many biological contexts such as spindles, in the cell cortex, and in acto-myosin rings. In these same contexts filament nucleation is tightly orchestrated by nucleator molecules. Many of these---TPX2 in spindles \cite{king2020phase} or WASP in cortices \cite{yan2022condensate} ---have been argued to phase separate. Intriguingly, damped oscillations, reminiscent of the ones from our model, have also been observed during actomyosin cortex in \emph{C. elegans} oocytes \cite{yan2022condensate}. While the detailed interactions between actin and WASP are likely more complex than in our model, what we present here provides a first pass at the problem. 

More broadly, length scale selection and memory fixation are important problems in cell biology. Phase separating systems with active chemistry can select length scales \cite{zwicker2015suppression, brauns2021wavelength, weyer2023coarsening}.
In chemically active droplets, growth rates around the trivial steady-state are altered by active processes and coarsening is linearly suppressed \cite{weyer2023coarsening, frohoff2021suppression}. In contrast, in our mechanically active droplets, coarsening suppression is non-linear, which is more similar to the non-reciprocal Cahn-Hilliard (NRCH) model \cite{saha2020scalar, frohoff2021suppression,frohoff2023non}.
Further, unlike in chemically active droplets, the phase separating quantity of our model stays conserved. Saving this conservation relation requires introducing new fields. In NRCH systems, this second field is another CH phase separating quantity, while in our model it is $\Phi$. In our model the mass of the nucleator and the momentum of the compount material are conserved.

Finally, in our system, the wavenumber of the final state is not independent of the initial condition. A broad spectrum of patterned states (oscillatory or stable) with steady wavenumbers are supported. This enables retaining information about the initial state, and the history of the system.

Together, our findings thus establish the compound material made of a phase separating nucleator and a self-straining network as a biologically plausible paradigm for mechano-chemical scale selection and memory fixation in cells.

\begin{acknowledgments}
This project has been funded by the Vienna Science and Technology Fund (WWTF) [10.47379/VRG20002].  Calculations were performed using supercomputer resources provided by the Austrian Scientific Computing Infrastructure (ASC). 
\end{acknowledgments}

\appendix

\section{Derivation of the equations of the cytoskeletal network}
\label{app:A}

In this section we derive the equations of motion for the self straining cytoskeletal network. In this paper we are concerned with the the 1D case, a more general discussion can be found in \cite{furthauer2019self, furthauer2021design}.  Our derivation starts from a phenomenological model for filament-filament interactions (Sec. \ref{app:A.1}), then solves the force balance of filaments in order to determine their velocities (Sec. \ref{app:A.2}). In (Sec. \ref{app:A.3}) we coarse grain the results from (Secs. \ref{app:A.1} and \ref{app:A.2}) to arrive at the equations of motion introduced in Sec. \ref{sec:eq}. 

\subsection{Crosslink-mediated forces between cytoskeletal filaments}
\label{app:A.1}

\begin{figure}[htbp]
    \centering
    \includegraphics[width=0.8\linewidth]{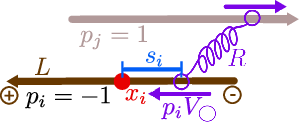}
    \caption{Sketch of crosslink-mediated filament interactions}
    \label{fig:var_names}
\end{figure}

Consider a system made of a large number of cytoskeletal filaments with center of mass positions $\mathbf x_i$, orientations $\mathbf p_i$, and lengths $L_i$. In 1D the unit vector $\mathbf p_i = \pm \hat e_x$. Here, $i$ is a particle index and $\hat e_x$ is the unit vector pointing in the positive $x$ direction.  Nearby filaments will be crosslinked by molecular scale motors, which exert forces between the points $s_i$, $s_j$ that they connect. Here $s_i, s_j$ denote the distances from the center of masses at which the crosslink is attached.  The sketch in Fig. \ref{fig:var_names} clarifies the geometry of the problem. 

In our model each crosslink exerts a force $f_{ij}$ between the filaments that it connects. Following \cite{furthauer2019self,furthauer2021design} we will postulate that 
\begin{equation}
f_{ij} =  -\gamma (v_i + V_{||} p_i - v_j - V_{||} p_j),
\end{equation}
where $\gamma$ is a friction coefficient and $V_{||}$ is the unloaded velocity of a motor head. Further $v_i = \partial_t {x_i}$ is the velocity if the center of mass of filament $i$.
The total force $F_i$ on filament $i$ is given by
\begin{widetext}
\begin{equation}
F_{i} = \sum_j\int_{-\frac{L_i}{2}}^{\frac{L_i}{2}} \diff s_i \int_{-\frac{L_j}{2}}^{\frac{L_j}{2}} \diff s_j \int_{\Omega(x_i + s_i p_i)}  \beta(s_i,s_j) \delta(y - x_j - s_j p_j)f_{ij} \diff y, 
\end{equation}
\end{widetext}
where $\beta$ quantifies the concentration of bound crosslinks; see \cite{furthauer2021design} for generalizations. Finally $\Omega$ encodes for the domain over which s crosslinker can connect two filaments. 

\subsection{Force balance and filament motion}
\label{app:A.2}

In the overdamped limit, the equation of motion for filament $i$ is simply $F_i=0$. This is trivially fulfilled when $v_i = - V_{||}p_i +v_0$ for all filaments, since then $f_{ij} = 0$ for all $i,j$. The constant velocity $v_0$ is fixed by requiring that the center of mass of the system be at rest. In our case $v_0 = 0$, since we consider a system in which equal filament masses point in either direction at any given time. Thus, filaments in the self straining network simply obey
\begin{equation}
v_i = -V_{||} p_i.
\label{eqs:vi}
\end{equation}

\subsection{Continuous fields and fluxes}
\label{app:A.3}

To relate the above force balance and equations of motion to the continuous equations of motion in the main text, we define the continuous density $\rho$, polarity $P$ and velocity $v$ as
\begin{subequations}
\begin{eqnarray}
    \rho(x) &=& \sum_i\delta(x_i-x),\\
    P(x) &=& \frac{1}{\rho}\sum_i p_i\delta(x_i-x),\\
    v(x) &=& \frac{1}{\rho}\sum_i v_i\delta(x_i-x)\label{eqs:vc},
\end{eqnarray}
\end{subequations}
respectively.
Using Eq. (\ref{eqs:vi}) in 
Eq. (\ref{eqs:vc}) immediately gives $v(x) = -V_{||}P(x)$. In the self straining network each filament moves in the direction set by its orientation at the preferred velocity of the motor.

The flux terms in Eqs. (\ref{eq:cyto}) can now be obtained by a simple time derivative
\begin{widetext}
\begin{subequations}
\begin{eqnarray}
    \partial_t \rho(x) &=& -\partial_x \sum_i v_i\delta(x_i-x) + S_\rho = -\partial_x (v\rho) + S_\rho = V_{||}\partial_x (P\rho) + S_\rho,\\
    \partial_t [\rho P](x) &=& - \partial_x \sum_i p_i v_i \delta(x_i-x)  + S_P = V_{||}\partial_x \rho + S_P,
\end{eqnarray}
\end{subequations}
\end{widetext}
where we used that $p_i^2=1$ by definition. Further, we introduced the source terms $S_P$ and $S_\rho$. They capture the fact that filaments are nucleated and fall apart and, in our model, are given by
\begin{subequations}
\begin{eqnarray}
S_\rho &=& -r \rho + \alpha_0 + \alpha c\rho, \\
S_P &=& -r \rho P. 
\end{eqnarray}
\end{subequations}
This modeling choice describes branching nucleation, with no preferred directionality.

\section{Deriving $q$-periodic (unstable) Cahn-Hilliard steady-state profiles and their harmonic amplitudes}
\label{app:B}

In the work presented in Sec. \ref{sec:scales}, we make use of an approximation which expresses the harmonics of the profile of a steady solution to the CH equation in terms of the wavenumber $q$, the mean density $c_0$ and the first harmonic $c_1$ (see Eq. (\ref{eq:tilde})). In this section we discuss the derivation and validity of this assumption. For this we first (Sec. \ref{app:B.1}) introduce an analytical expression for the steady CH profiles exploiting the fact that at steady state the chemical potential is spatially uniform. We next (Sec. \ref{app:B.2}) introduce a series of harmonic amplitude equations, which underlie the dimensional reduction that we aim for. In this we will use the results of (Sec. \ref{app:B.1}) to formulate (and validate) the closure assumption, which enables us to do this.

\subsection{Profile derivation using the chemical potential}
\label{app:B.1}

\begin{figure}[t]
\centering
\includegraphics[width=\linewidth]{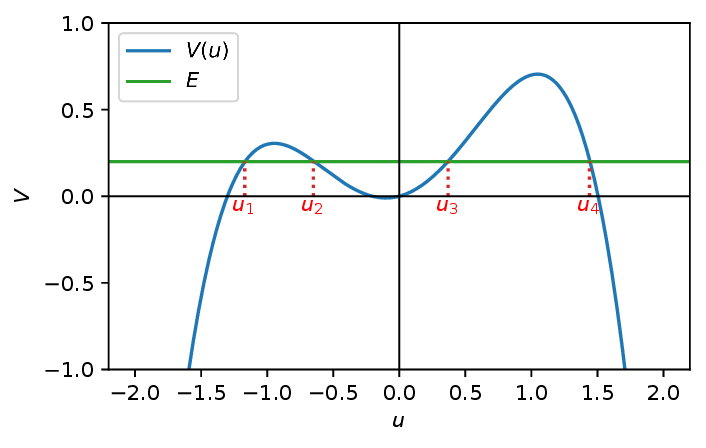}
\caption{Blue: $V(u)$ as a function of $u$, for unitary parameters ($a=b=\kappa=1$), $\mu = 0.1$ and $E=0.2$. $E$ is plotted in green.}
\label{fig:V}
\end{figure}

In this section, we derive stationary profiles of the Cahn-Hilliard equation
\begin{equation}
\label{eq:CH}
    \partial_t c = \Lambda \partial_x^2\left(  a(c-c_\mathrm{c})^3 - b(c-c_\mathrm{c}) - \kappa \partial_x^2 c)\right).
\end{equation}
Our approach closely follows the method of Argentina \emph{et al.} \cite{argentina2005coarsening}. Accordingly, we will use $u = c - c_\mathrm{c}$, and fall in line with the notations of \cite{argentina2005coarsening}. In this notation the chemical potential $\mu$ reads
\begin{equation}
\label{eq:mu}
\mu = a u^3 - b u - \kappa \partial_x^2 u.
\end{equation}
In steady states, $\mu$ is spatially uniform
as we assumed that there is no source or sinks of $c$ in our system \cite{langer1971theory}. Rewriting the last equation as $\kappa \partial_x^2 u = b u - au^3 +\mu$, one notices the formal resemblance with Newton's second law, usually expressed for $(x,t)$ instead of $(u,x)$. Following up on this idea, we will get to an energetic formulation, multiplying by $\partial_x u$ and integrating:
\begin{equation}
2\frac{\mu}{\kappa} u - \frac a{2 \kappa} u^4 + \frac{b}{\kappa} u^2 + (\partial_x u)^2 = E,
\end{equation}
with $E$ a constant that can be likened to energy in our analogy. From this, it is possible to define a potential (Fig. \ref{fig:V})
\begin{equation}
\label{eq:V}
V(u) = 2\frac{\mu}{\kappa} u + \frac{b}{\kappa} u^2 - \frac{a}{2 \kappa} u^4,
\end{equation}
so that one eventually has
\begin{equation}
(\partial_x u)^2 + V(u) = E.
\end{equation}

Therefore, getting the spatial profile of $u$ boils down to solving the time evolution of a particle in the potential $V$. In our case of interest, $E - V(u)$ has four roots (there might be identical roots), and in our analogy, our particle is confined in the central well. This yields
\begin{equation}
\label{eq:E-V_1}
(\partial_x u)^2 = E - V(u) = \frac{a}{2\kappa}(u-u_1)(u-u_2)(u-u_3)(u-u_4),
\end{equation}
with $u_1 \leq u_2 \leq u_3 \leq u_4$. The $\frac{a}{2\kappa}$ is arbitrary, but will be essential in the following. Finally, it is possible to integrate this relation, and to get the following expression for the space profile of $u$:
\begin{widetext}
\begin{equation}
u(x) = u_4 - \frac{(u_4-u_3)(u_4-u_2)}{(u_4-u_2) + (u_3-u_2) ~ \mathrm{sn}^2 \left(\sqrt{\frac{(u_3-u_1)(u_4-u_2)}4}\sqrt{\frac{a}{2\kappa}}x,\frac{(u_4-u_1)(u_3-u_2)}{(u_3-u_1)(u_4-u_2)}\right)},
\end{equation}
\end{widetext}
where $\mathrm{sn}$ is a Jacobi elliptic function \cite{bowman1953introduction}. Details of this integration can be found in \cite{novick1984nonlinear, argentina2005coarsening}.

In the following, we will make approximations to relate $u_1$, $u_2$, $u_3$ and $u_4$ back to our physical parameters. Note that we can immediately link them to $E$, $\mu$, $b/a$ and $\kappa/a$ using Eq. (\ref{eq:V}). However, $E$ and $\mu$ are not part of our physical parameters and we will preferably use $c_0$ (or $u_0 = c_0 - c_\mathrm{c}$) and $q$ (the wavenumber of the profile).
This requires performing integrals (as $c_0$ is the average of $c$), and we will thus need to convert the $\mathrm{sn}$ function into a hyperbolic tangent, i.e. to assume that the space period of the pattern is infinite. Note that for Cahn-Hilliard profiles, the spatial length of the border in between two plateaus (i.e. the typical size of a droplet) is $(\frac{b}{2\kappa})^{1/2}$, while the fastest growing mode has a period $2\pi(\frac{b}{2\kappa})^{1/2}$. The infinite wavelength assumption is thus reasonable. We now start to eliminate the $u_i$ within this approximation.

The Jacobi elliptic function has an infinite period if $(u_4-u_1)(u_3-u_2) = (u_3-u_1)(u_4-u_2)$, which yields $u_1=u_2$ or $u_3=u_4$. If $\mu>0$, $u_3=u_4$ would lead to an unbounded $c$-profile, which is unphysical. In contrast the condition $u_1 =u_2$ produces a droplet structure in which $u_3$ is the value of $u$ in the middle of the droplet and $u_1<0$ is the value of $u$ far away from the droplet. Note that the case of $\mu <0$ can be treated in strict analogy, by redefining Eq.~\ref{eq:mu} in terms of $\tilde u = -u$. 

Still using $\mu>0$ and thus $u_1=u_2$ we can now link $u_1$ to $\mu$ by requiring $\partial_u V (u_1) = 0$, and obtain
\begin{equation}
\mu = - b u_1 + a u_1^3,
\end{equation}
and thus,
\begin{eqnarray}
\nonumber E = V(u_1) & = & \frac{2}{\kappa}\left(-b u_1+a u_1^3\right) u_1 + \frac{b}{\kappa} u_1^2 - \frac{a}{2\kappa} u_1^4 \\
\label{eq:E-V_2} & = & -\frac{b}{\kappa} u_1^2 + \frac{3a}{2} u_1^4.
\end{eqnarray}
Using Eq. (\ref{eq:V}), Eq. (\ref{eq:E-V_2}) eventually yields an expression for $E - V(u)$ that we can compare with Eq. (\ref{eq:E-V_1}):
\begin{equation}
    E-V(u) = \frac{a}{2\kappa} (u-u_1)^2(u+u_1+\delta)(u+u_1-\delta),
\end{equation}
where we introduced 
\begin{equation}
\label{eq:def_delta}
\delta = \sqrt{2 \frac{b}{a} - 2u_1^2}.
\end{equation}
This way, our problem now only depends on $\kappa/a$ and two among $b/a$, $\delta$ and $u_1$. In the following, we will get back to $q$-periodic profiles of average $c_0$, and thus replace $\delta$ (and the redundant $u_1$) by $q$ and $c_0$.

\begin{figure*}[t]
\centering
\includegraphics[width=\linewidth]{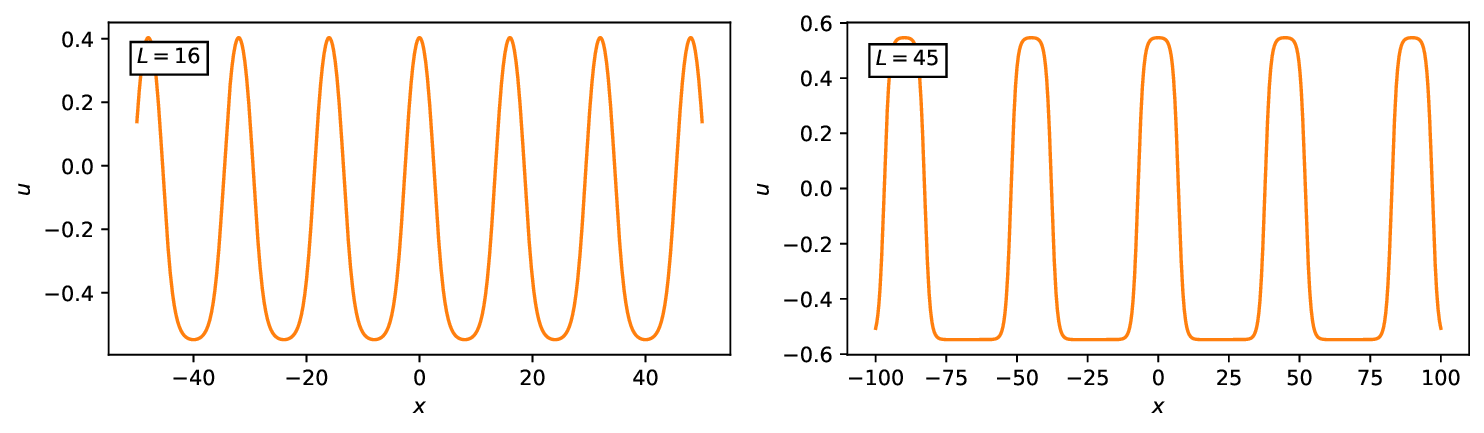}
\caption{Different profiles computed using Eq. (\ref{eq:profile}) for the parameter values introduced in Fig. \ref{fig:simu}: $u_0=-0.2$, $b/a=0.3$, $\kappa/a=0.525$. Left: $L=16$; Right: $L=45$.}
\label{fig:profile}
\end{figure*}

Considering that the period is infinite (i.e. $u_1 = u_2$) also implies that the Jacobi elliptic function $\mathrm{sn}$ has converged to a hyperbolic tangent (its modulus is $1$). Putting all this together, one gets
\begin{equation}
    u(x) = \delta -u_1-\frac{2\delta(\delta-2u_1)}{\delta-2u_1 - (\delta + 2u_1) \tanh^2\left(\sqrt{\frac{3a u_1^2 - b}{\kappa}} x \right)}.
\end{equation}
Note that we keep using both $u_1$ and $\delta$ for simplicity, even if everything may be written with $\delta$, $b/a$ and $\kappa/a$.

However, when the period grows, $u_1$ converges toward $-\sqrt{b/a}$ which is the lower plateau value of Cahn-Hilliard final profile \cite{weber2019physics}. Therefore, $\delta$ converges towards zero, and it is more convenient to rewrite this formula with an hyperbolic cosine:
\begin{eqnarray}
\nonumber u(x) & = & u_1 - \frac{(\delta - 2u_1)(-\delta - 2u_1)}{-2u_1 - \delta \cosh\left(\sqrt{\frac{3a u_1^2 - b}{\kappa} x}\right)}\\
\label{eq:cosh_1} & = & u_1 - \frac{6(u_1^2 - \frac{b}{3a})}{-2u_1 - \delta \cosh\left(\sqrt{\frac{3a u_1^2 - b}{\kappa} x}\right)}.
\end{eqnarray}
This result is known, and has already been derived in \cite{novick1984nonlinear, langer1971theory, argentina2005coarsening}. Note that we exactly followed the method depicted in \cite{argentina2005coarsening} to obtain Eq. (\ref{eq:cosh_1}). In the following, we use this result to perform integrals, and get back to a periodic profile, rather than a single bubble.

Indeed it is clear, as we supposed an infinite period, that the spatial average of $c$ is no longer $c_0$. Enforcing this condition will enable us to know where to cut the infinite $u=u_1$ tails of this profile, in order to build a periodic solution. Integrating Eq. (\ref{eq:cosh_1}) as in \cite{argentina2005coarsening} and eliminating the $u_1$ yields (with $L=2\pi/q$)
\begin{eqnarray}
\nonumber c_0 L & = & u_1 L + \sqrt{\frac{8\kappa}a} ~ \mathrm{arccosh}\left(\frac{-2u_1}{\delta}\right)\\
\nonumber & = & \sqrt{\frac{b}{a}-\frac{\delta^2}{2}} L + \sqrt{\frac{8\kappa}a} ~ \mathrm{arccosh}\left(\frac{2\sqrt{b/a - \delta^2/2}}{\delta}\right),\\
\label{eq:delta_1}
\end{eqnarray}
which directly links the space period $L$ and $\delta$. Supposing that $\delta^2 \ll 2 b/a$ (i.e. $u_1 \to -\sqrt{b/a}$), one eventually gets
\begin{equation}
\label{eq:delta_2}\delta = \frac{2\sqrt{b/a}}{\cosh\left(\frac{u_0+\sqrt{b/a}}{\sqrt{8\kappa /a}}L\right)}.
\end{equation}
Please note that Eq. (\ref{eq:delta_2}) is an approximation supposing $\delta^2 \ll 2b/a$. Eq. (\ref{eq:delta_1}) may be used for better precision.

\begin{figure*}[t]
    \centering
    \includegraphics[width=\linewidth]{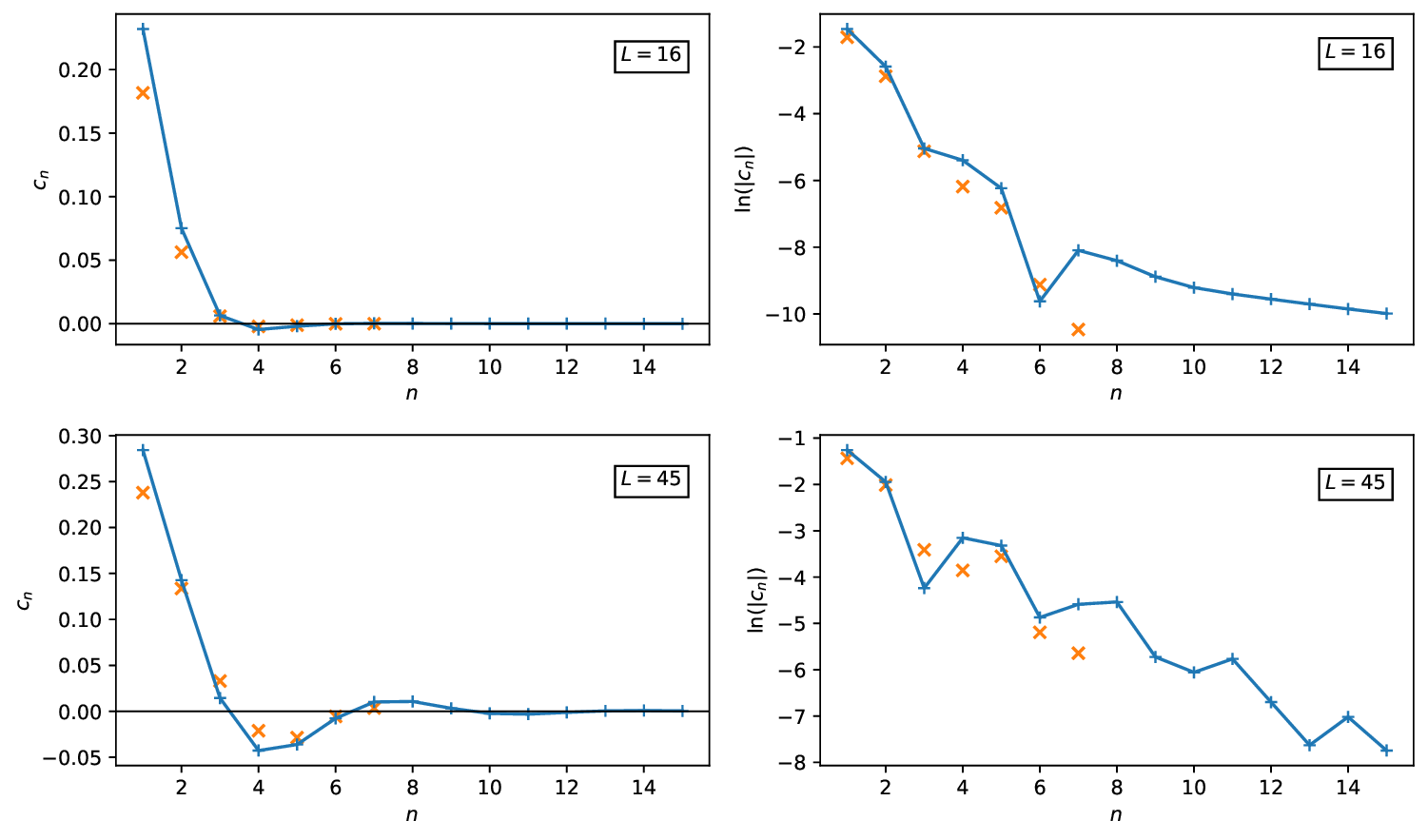}
    \caption{Blue: values of the Fourier coefficients of the profile from Eq. (\ref{eq:profile}); Orange: sequential computation of the harmonics using truncated Eq. (\ref{eq:c_poly}). Top: $L=16$; Bottom: $L=45$. The other coefficients are the same as in Fig. \ref{fig:profile}. Left: linear scale. Right: log-lin scale.}
    \label{fig:harmonics}
\end{figure*}

Finally, with given $L$ and $c$ we can compute $\delta$, and it is therefore possible to build the modulus $m$ of $\mathrm{sn}$ such that the space period is correct. The complete elliptic integral of the first kind $K$ has to verify
\begin{eqnarray}
\nonumber K(m) &:=& \int_0^1 \frac{\mathrm dt}{\sqrt{(1-t^2)(1-mt^2)}}\\
\label{eq:m} &=& \frac{\sqrt{3au_1^2-b}}{4\sqrt{\kappa}}L = \frac{\sqrt{2b - 3a\frac{\delta^2}{2}}}{4\sqrt{\kappa}}L.
\end{eqnarray}
Therefore, we can eventually give the final expression of our approximation of $u$ (see Fig. \ref{fig:profile}):
\begin{widetext}
\begin{equation}
\label{eq:profile}
u(x) = \sqrt{\frac{b}{a} - \frac{\delta^2}{2}} + \delta - \frac{4\sqrt{\frac{b}{a} - \frac{\delta^2}{2}}\delta + 2 \delta^2}{2\sqrt{\frac{b}{a} - \frac{\delta^2}{2}} + \delta - \left(2\sqrt{\frac{b}{a} - \frac{\delta^2}{2}} - \delta\right) ~ \mathrm{sn}^2\left(\frac{\sqrt{2 b - 3a\frac{\delta^2}{2}}}{2\sqrt{\kappa}}x, m\right)}.
\end{equation}
\end{widetext}
This profile depends on the parameters $\frac{b}{a}$, $\frac{\kappa}{a}$, $q$ and $u_0$. The two last ones parametrize one specific stationary solution, and are used in the calculation of $\delta$ in Eq. (\ref{eq:delta_1}) (or Eq. (\ref{eq:delta_2}) for a simplified expression) and of $m$ in Eq. (\ref{eq:m}).

\subsection{Harmonic amplitude equations}
\label{app:B.2}
\label{sec:hae}
In this section, we use the Fourier decomposition of the $c$ profile to compute the stationary Fourier coefficients (i.e. the Fourier coefficients of Eq. (\ref{eq:profile})).
We define the Fourier coefficients by
\begin{equation}
\label{eq:c_Fourier}
c(x) = c_0 + 2 \sum_{n\geq1} c_n \cos(nqx),
\end{equation}
where all $c_n$ are real, since we chose the coordinate system such that the profile is even.
Thus, we consider profiles with period $L$, such that $L=\frac{2\pi}q$, where $q$ is the wavenumber. The equation of evolution of $c$ yields a hierarchy of equations for the Fourier amplitudes, which reads (defining $c_{-n} = c_n$)
\begin{widetext}
\begin{eqnarray}
\label{eq:hae}
    \partial_t c_n &= - n^2 q^2 \Lambda \Bigg( & \Big( 3a(c_0-c_\text{c})^2 - b + n^2 q^2 \kappa \Big) c_n + 3a(c_0-c_\text{c}) \underset{i+j=n}{\sum_{(i,j) \in {\mathbb{Z}^*}^2}} (c_i c_j) + a \underset{i+j+k=n}{\sum_{(i,j,k) \in {\mathbb{Z}^*}^3}} (c_i c_j c_k) \Bigg) \\
\nonumber
    &= - n^2 q^2 \Lambda \Bigg( & 3 a c_n^3 + 3 a c_{3n}c_n^2 + \Big( 3 a (c_0-c_\text{c})^2 - b + n^2 q^2 \kappa + 6a \underset{i\neq n}{\sum_{i\geq1}} c_i^2 + 3a \underset{i+j = 2n}{\sum_{(i,j) \in (\mathbb{Z}\setminus\{0,\pm n\})^2}} (c_i c_j) \Big) c_n \\
\label{eq:c_poly}
    && + \Big( 3 a (c_0-c_\text{c}) \underset{i+j = n}{\sum_{(i,j) \in (\mathbb{Z}\setminus\{0, \pm n\})^2}} (c_i c_j) + a \underset{i+j+k = n}{\sum_{(i,j,k) \in (\mathbb{Z}\setminus\{0, \pm n\})^3}} (c_i c_j c_k) \Big) \Bigg),
\end{eqnarray}
\end{widetext}
where, in Eq. (\ref{eq:c_poly}), we casted the right hand side as a third degree polynomial in $c_n$.
This set of equations couples all $c_n$. Thus, in practical terms, to solve for steady states, a closure approximation is needed to truncate the hierarchy.

To formulate a closure relation we use
the analytic steady state profiles derived in section II.A.
We find that $|c_n|$ can be considered---at least for the first ones---as decaying as $\exp(-n)$ (see Fig. \ref{fig:harmonics}). This will be the basis for our closure approximation.
Note that later in the coarsening process, our closure approximation becomes progressively worse.
In the following, we take $c_i \sim \varepsilon^i$, where $\varepsilon \ll 1$. Therefore, for $c_1$, if
we truncate the coefficients of polynomial order $\varepsilon^2$, we obtain the following expression :
\begin{equation}
\label{eq:dtc1}
    \partial_t c_1 = -q^2 \Lambda \left( 3a c_1^3 + \left(3a(c_0-c_\text{c})^2 - b + q^2\kappa\right) c_1 \right).
\end{equation}
Note, $c_3\sim\varepsilon^3$ is neglected, and the second degree coefficient vanish; the first sum of the first degree coefficient contains exponentially decaying terms of order larger that $\varepsilon^4$ (the first term being $c_2^2\sim\varepsilon^4$), as does the second sum, with term orders larger that $\varepsilon^5$ (the first term being $c_2c_3\sim\varepsilon^5$), while $3a(c_0-c_\text{c})^2 - b + n^2 q^2\kappa \gtrsim \varepsilon^2$; and the zero order coefficient contains exponentially decaying terms of order larger than $\varepsilon^6$ (note that $c_0-c_\text{c} \sim \varepsilon$ for the set of parameters we consider).

\begin{figure}[t]
\centering
\includegraphics[width=\linewidth]{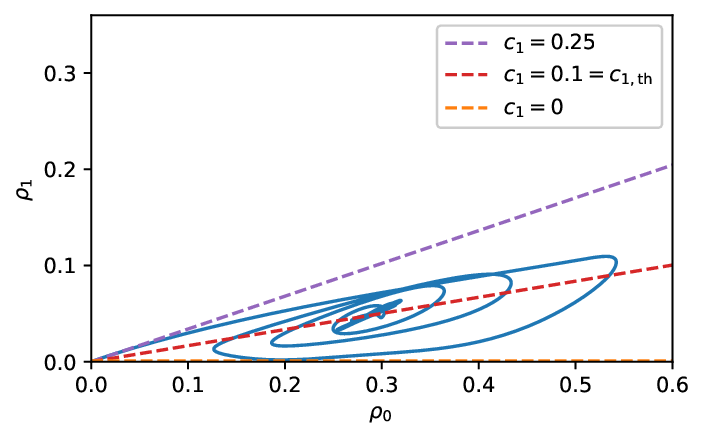}
\caption{Trajectory of the simulation depicted in Fig. \ref{fig:simu} in the plane $(\rho_0,\rho_1)$. In dashed lines, $\vec\rho_{\sigma}$ for $c_1=0.25$, $c_1=0.1=c_{1,\text{th}}$ and $c_1=0$, in purple, red and orange, respectively, computed with $M_1$.}
\label{fig:vec_rho}
\end{figure}

Therefore, we obtain an approximation giving $c_{1,\text{CH}}$ only as a function of the parameters of the system. For $n\leq3$, a similar truncation at $\varepsilon^{2n}$ gives approximations $c_{n,\text{CH}}(c_1,\dots c_{n-1})$, which enables to compute the harmonics sequentially. For larger $n$, truncating to $\varepsilon^{2n}$ no longer provide this approximation. For example, for $n=5$, the last sum contains a term $c_1^2 c_7 \sim \varepsilon^9$. Furthermore, the second sum of the first degree coefficient contains a linearly growing number of $\varepsilon^{2n}$ coefficients ($c_1c_9$, $c_2c_8$,\dots , for $n=5$), which are to numerous to be neglected for a large $n$. In this case, the truncation has to be adapted, and the error made in approximating $c_{n,\text{CH}}(c_1,\dots c_{n-1})$ is increasing. Nevertheless, considering the first few harmonics will be sufficient in our study. For the next two harmonics, one obtains
\begin{widetext}
\begin{subequations}
\label{eq:trunc}
\begin{eqnarray}
\frac{\partial_t c_2}{-4q^2\Lambda} & = & 3a c_2^3 + \left(3a(c_0-c_\mathrm{c})^2- b + 4q^2 \kappa + 6a c_1^2 \right) c_2 + 3a(c_0-c_\mathrm{c}) c_1^2, \\
\frac{\partial_t c_3}{-9q^2\Lambda} & = & 3 a c_3^3 + \left(3a(c_0-c_\mathrm{c})^2- b + 9q^2\kappa + a (6 c_1^2 + 6 c_2^2) \right) c_3 + 3a(c_0-c_\mathrm{c}) \left(2c_1c_2\right) + a \left(c_1^3 + 3c_1c_2^2 \right).
\end{eqnarray}
\end{subequations}
\end{widetext}

Fig. \ref{fig:harmonics} shows the sequential computation of these harmonics (up to $c_7$). Note that $c_{n\geq4}$ are computed using truncation at orders lower than $\varepsilon^{2n}$, in order to obtain an approximation depending only on the previous harmonics. The result shows a relatively good agreement as long as $c_n\sim\varepsilon^n$. For the study we will perform, these approximations will be sufficient. Furthermore, Eq. (\ref{eq:trunc}) shows that the dynamics of the harmonics $c_n$ are getting faster and faster when $n$ grows (due to the $n^2$ terms).

\section{Deriving the threshold $c_{1, \text{\rm th}}$, beyond which $\rho$ grows}
\label{app:C}

We now seek the threshold on the $c$ profiles for $\rho$ to depart from its trivial profile. Let us decompose $\rho$ and $\Phi$ in a similar way as we did for $c$ in Eq. (\ref{eq:c_Fourier}). One obtains
\begin{subequations}
\begin{eqnarray}
\rho (x) & = & \rho_0 + \left(\rho_1 \mathrm{e}^{\mathrm{i}qx} + \rho_2 \mathrm{e}^{\mathrm{i}2qx} + \dots + \mathrm{c.c.} \right), \\
\Phi (x) & = & \Phi_0 + \left( \Phi_1 \mathrm{e}^{\mathrm{i}qx} + \Phi_2 \mathrm{e}^{\mathrm{i}2qx} + \dots + \mathrm{c.c.} \right).
\end{eqnarray}
\end{subequations}
We can now write the linear system obtained by injecting this decomposition the equations of the cytoskeletal network, which yields Eqs. (\ref{eq:hierarchy}).
Note that to obtain it, we took $\rho_n$ real, assuming that $c$ and $\rho$ are in phase. This is motivated by our numerical simulations, which show this to be the case close to the threshold.

It is instructive to first discuss an approximation where only the first harmonics of $\rho$ and $\Phi$ are considered. In this case, $M$ is only a $4\times4$ matrix. Furthermore, it is even possible to drop the $P_0$ variable, as we saw it is coupled with no other variable, and vanishes. Therefore, we get the following matrix
\begin{equation}
M_1 = 
\begin{array}{c}
\rho_0 \\ \rho_1 \\ P_1
\end{array}
\left(\begin{matrix}
\alpha c_0 - r & 2 \alpha c_1 & 0 \\
\alpha c_1 & \alpha c_0 - r &  \mathrm{i}qv \\
0 & \mathrm{i}qv & -r
\end{matrix}\right)
\end{equation}
by also truncating the $c$ harmonics to $c_1$.
We can obtain the condition on $c_1$ to have $\det(M_1) = 0$ (i.e. $0$ is an eigenvalue). It reads
\begin{equation}
c_{1,\text{th}} = \frac1{2\alpha}\sqrt{2(r-\alpha c_0)\left(r-\alpha c_0 + \frac{q^2v^2}{r}\right)}.
\end{equation}
Importantly, this expression already shows that the threshold depends on both the amplitude and the wavenumber of the $c$-profile.

More generally, we can do the very same thing with the full matrix $M$, providing us a condition on the $\{c_n\}$ families for $M$ to have a positive eigenvalue. In order to obtain a threshold on $c_1$, we can reduce the dimension of the space of $\{c_n\}$ families by using Eq. (\ref{eq:tilde}). Indeed, for any $c_1$, we can compute a family of steady $c_{n\geq2}$. Thus, we only consider a 1-dimensional set of $\{c_n\}$ families, which includes the $\{c_{n,\text{CH}}\}$ family (it is the special case $c_1=c_{1,\text{CH}}$).

Furthermore, to probe the evolution of the whole $\{\rho_n\}$ family, one should look along the $\vec\rho_\sigma$ eigenvector, associated with the largest eigenvalue $\sigma$ of the system. Note that each $\sigma$ corresponds to a value of $c_1$, for instance, $\sigma(c_1=c_{1,\text{th}})=0$. Moreover, the $\{\rho_0, 0, \dots \}$ family is almost parallel to $\vec\rho_\sigma$ (see Fig. \ref{fig:vec_rho}). In the following, we will therefore use $\rho_0$ to probe the evolution of the $\{\rho_n\}$ family.

\begin{figure}[t]
\centering
\includegraphics[width=\linewidth]{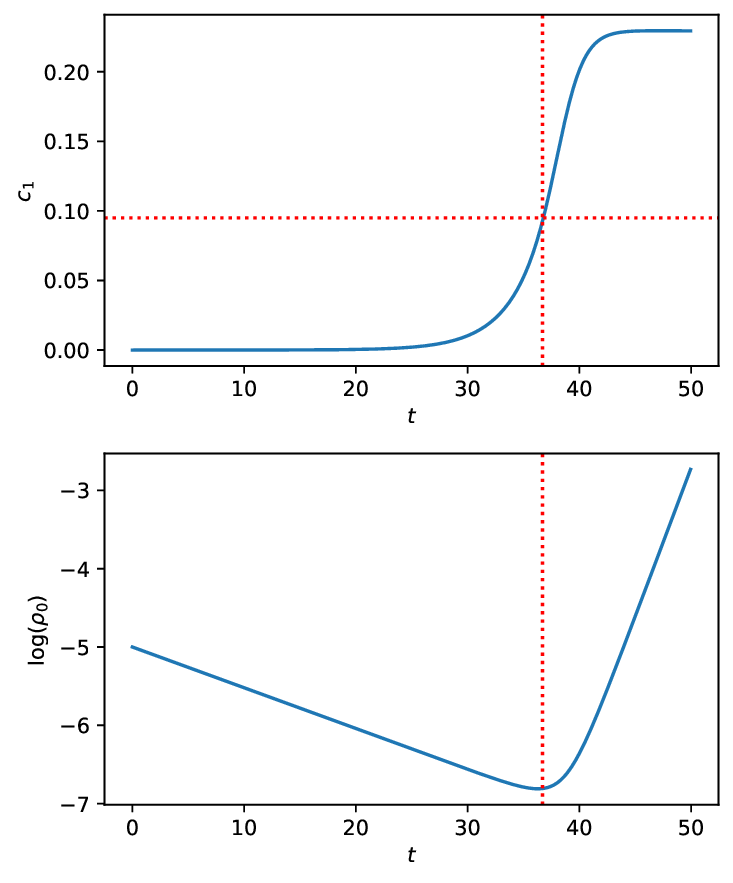}
\caption{Above: $c_1$ with respect to time; Below: $\log(\rho_0)$ with respect to time. We plotted the threshold obtained with Eq. (\ref{eq:hierarchy}): $c_{1,\text{th}}=0.095$. The parameter values are ones introduced in Fig. \ref{fig:simu}, except $\alpha_0=0$ and $\rho_\text{init}=10^{-5}$. Setting $\alpha_0=0$ helps seeing the threshold for the growth of $\rho_0$, without affecting the analysis we conducted.}
\label{fig:c1th_plot}
\end{figure}

Fig. \ref{fig:c1th_plot} shows the initial growth of $\rho_0$ happening for $c_1=0.095$. This is the threshold value computed with the matrix $M$. Note that $M_1$ yields $c_{1,\text{th}}=0.10$, and $M_2$ already yields $c_{1,\text{th}}=0.095$. Further harmonics only give small corrections.

\section{Deriving the threshold $\rho_{0,th}$ beyond which the dynamics of $c$ departs from CH}
\label{app:D}
\subsection{Adiabatic approximation}
\label{app:D.1}

In order to simplify the expressions, and to reduce the dimension of the space of profiles we consider, we will perform an adiabatic approximation to link $\rho$ and $\Phi$. On the onset of the growth/decay of $\rho$, $\Phi$ evolves with a much faster rate than $\rho$, hence the following approximation:
\begin{equation}
    \label{eq:adiabatic}
    \Phi = \frac{V_{||}}{r} \partial_x \rho.
\end{equation}

\begin{figure}[t]
    \centering
    \includegraphics[width=\linewidth]{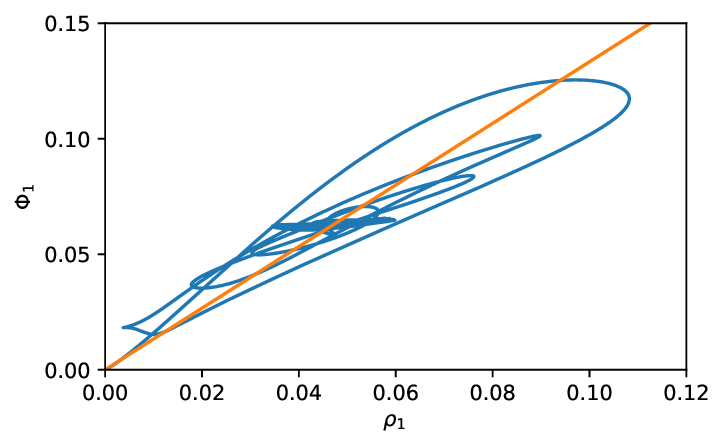}
    \caption{Blue: Trajectory of the simulation depicted in Fig. \ref{fig:simu} in the plane $(\rho_1,\Phi_1)$. Orange: Proportionality relation given by the approximation Eq. (\ref{eq:adiabatic}).}
    \label{fig:adiabatic}
\end{figure}

This approximation is justified on the onset of the growth/decay of $\rho$, and can therefore be used to simplify the matrix $M$ when looking for zero eigenvalues. We will however use it throughout the dynamics of the system. Indeed, simulation shows that it remains a reasonable assumptions even further from the onset. Fig. \ref{fig:adiabatic} support this approximation.

\begin{figure}[t]
    \centering
    \includegraphics[width=\linewidth]{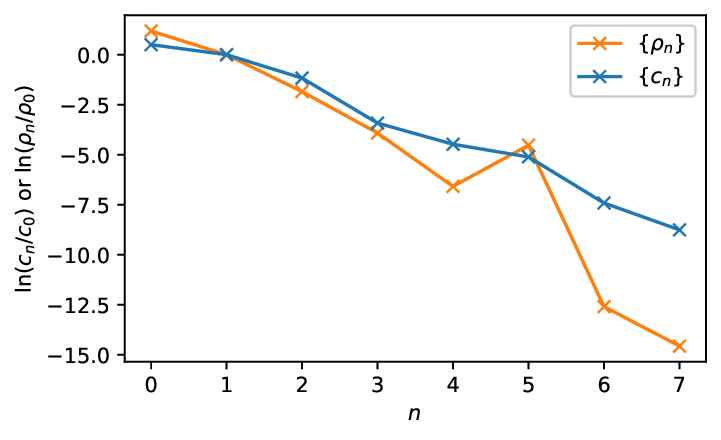}
    \caption{Harmonics of $\rho$ and $c$ in a log-lin scale, computed using $\vec\rho_\sigma$ and truncated Eq. (\ref{eq:c_poly}). Note that the vector $\vec\rho_\sigma$ and the family $\{c_n\}$ are normalized such that $c_1 = \rho_1$.}
    \label{fig:rho_harmonics}
\bigskip
    \includegraphics[width=\linewidth]{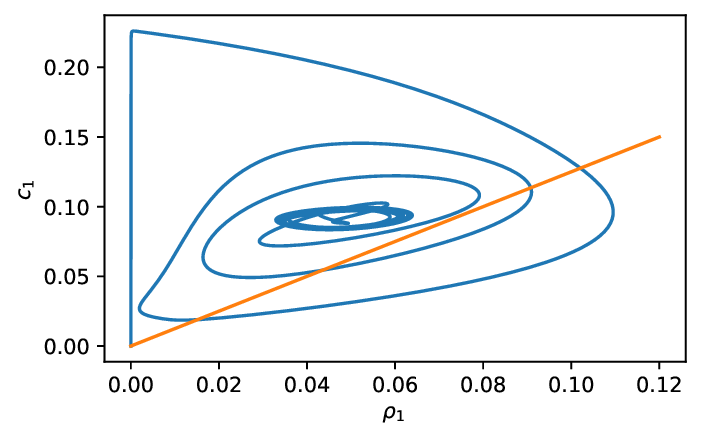}
    \caption{Blue: Trajectory of the simulation depicted in Fig. \ref{fig:simu} in the plane $(\rho_1,c_1)$. Orange: $c_1=\rho_1$ line. This shows that throughout the dynamics of the system, $\rho_1 \lesssim c_1$.}
    \label{fig:snail}
\end{figure}

\subsection{Harmonic development of $c$}
\label{app:D.2}

Let us reuse the harmonic amplitude equations of Appendix \ref{sec:hae}. However, now, $\rho$ and $\Phi$ are no longer null, and add a new term in Eq. (\ref{eq:hae}) and Eq. (\ref{eq:c_poly}). In Eq. (\ref{eq:hae}), this supplemental term reads
\begin{equation}
\label{eq:new_term}
    - kV_{||}n^2q^2 \sum_{i\in\mathbb{Z^*}} \rho_i c_{n-i}.
\end{equation}
In order to compare this term with the previous ones, and to perform a similar truncation as in Eq. (\ref{eq:trunc}), we can suppose that $\rho_n \lesssim \varepsilon^n$. This assumption is corroborated by Fig. \ref{fig:rho_harmonics} and Fig. \ref{fig:snail}.

When plugging Eq. (\ref{eq:new_term}) in Eq. (\ref{eq:c_poly}), and performing the previously depicted truncation, we remark that this term is non-negligible in Eq. (\ref{eq:dtc1}) for $n=1$. Indeed the leading term of Eq. (\ref{eq:new_term}), that is to say $\rho_1 c_0$, should not be neglected. For $n\geq2$, this sum is of same weight as the third sum of Eq. (\ref{eq:c_poly}). However,  with our set of parameters, the coefficient in front of it is much smaller. We can therefore neglect it, and keep Eq. (\ref{eq:trunc}) and the further equations unchanged.

\section{Stability of the steady states}
\label{app:E}

We now perform a linear stability analysis around the previously derived steady states. Let us consider a stationary state, and use only its sinusoidal approximation, i.e. truncate $\rho$, $c$, and $\Phi$ after the first harmonic. That means, we introduce
\begin{equation}
    X_\text{s}
    =
    \begin{pmatrix} c_0 \\ \rho_0 \\ 0 \end{pmatrix} + 
    \begin{pmatrix} c_q \\ \rho_q \\ \mathrm{i}\Phi_q \end{pmatrix}
    \left(\mathrm{e}^{\mathrm{i}qx} + \mathrm{e}^{-\mathrm{i}qx}\right)
\end{equation}
as our steady state of wavenumber $q$. Moreover, we add a small perturbation of wavenumber $p$, and also take into account the possible excitation of the modes $q-p$ and $q+p$ due to the non-linearities of the system. Eventually, we consider the following Ansatz:
\begin{eqnarray}
    \nonumber \begin{pmatrix} c \\ \rho \\ \Phi \end{pmatrix}
    &=&
    X_\text{s}
    +
    \begin{pmatrix} c_p \\ \rho_p \\ \mathrm{i}\Phi_p \end{pmatrix}
    \left(\mathrm{e}^{\mathrm{i}px} + \mathrm{e}^{-\mathrm{i}px}\right) \\
    \nonumber &&+
    \begin{pmatrix} c_{q+p} \\ \rho_{q+p} \\ \mathrm{i}\Phi_{q+p} \end{pmatrix}
    \left(\mathrm{e}^{\mathrm{i}(q+p)x} + \mathrm{e}^{-\mathrm{i}(q+p)x}\right)\\
    &&+
    \begin{pmatrix} c_{q-p} \\ \rho_{q-p} \\ \mathrm{i}\Phi_{q-p} \end{pmatrix}
    \left(\mathrm{e}^{\mathrm{i}(q-p)x} + \mathrm{e}^{-\mathrm{i}(q-p)x}\right).
\end{eqnarray}

It is possible to inject these expressions in our full system of equations, and to perform a linear stability analysis. We get the following matrix:
\begin{widetext}
\begin{equation}
\label{eq:big_matrix}
\begin{matrix}
    c_p \\ c_{q+p} \\ c_{q-p} \\ \rho_p \\ \rho_{q+p} \\ \rho_{q-p} \\ \Phi_p \\ \Phi_{q+p} \\ \Phi_{q-p}
\end{matrix}
\begin{pmatrix}
D(p) & 
N_+(p) & 
N_-(p) &
0 & 0 & 0 &
-p k v c_0 &
-p k v c_q &
 p k v c_q \\
N_-(q+p) & 
D(q+p) & 
0 &
0 & 0 & 0 &
-(q+p) k v c_0 &
 -(q+p) k v c_q &
0 \\
N_-(q-p) & 
0 & 
D(q-p) &
0 & 0 & 0 &
(q-p) k v c_0 &
0 &
-(q-p) k v c_q \\
\alpha \rho_0 & \alpha \rho_q & \alpha \rho_q &
\alpha c_0 - r & \alpha c_q & \alpha c_q &
-vp & 0 & 0 \\
\alpha \rho_q & \alpha \rho_0 & 0 &
\alpha c_q & \alpha c_0 - r & 0 &
0 & -v(q+p) & 0 \\
\alpha \rho_q & 0 & \alpha \rho_0 &
\alpha c_q & 0 & \alpha c_0 - r &
0 & 0 & -v(q-p) \\
0 & 0 & 0 & vp & 0 & 0 & -r & 0 & 0 \\
0 & 0 & 0 & 0 & v(q+p) & 0 & 0 & -r & 0 \\
0 & 0 & 0 & 0 & 0 & v(q-p) & 0 & 0 & -r \\
\end{pmatrix},
\end{equation}
\end{widetext}
where we defined the following expressions:
\begin{subequations}
\begin{eqnarray}
    D(p) &=& \Lambda p^2 \left(b-3a(c_0-c_\text{c})^2 - 6a c_q^2 - \kappa p^2\right), \\
    N_\pm (p) &=& -\Lambda p^2 6 a (c_0-c_\mathrm{c})c_q \pm kvp\Phi_q.
\end{eqnarray}
\end{subequations}

The same strategy can be applied for pure Cahn-Hilliard steady-states by considering $\rho = \Phi= 0$, therefore yielding a matrix similar to the upper left $3\times3$ block of Eq. (\ref{eq:big_matrix}).

\bibliography{references}

\begin{thebibliography}{47}%
\makeatletter
\providecommand \@ifxundefined [1]{%
 \@ifx{#1\undefined}
}%
\providecommand \@ifnum [1]{%
 \ifnum #1\expandafter \@firstoftwo
 \else \expandafter \@secondoftwo
 \fi
}%
\providecommand \@ifx [1]{%
 \ifx #1\expandafter \@firstoftwo
 \else \expandafter \@secondoftwo
 \fi
}%
\providecommand \natexlab [1]{#1}%
\providecommand \enquote  [1]{``#1''}%
\providecommand \bibnamefont  [1]{#1}%
\providecommand \bibfnamefont [1]{#1}%
\providecommand \citenamefont [1]{#1}%
\providecommand \href@noop [0]{\@secondoftwo}%
\providecommand \href [0]{\begingroup \@sanitize@url \@href}%
\providecommand \@href[1]{\@@startlink{#1}\@@href}%
\providecommand \@@href[1]{\endgroup#1\@@endlink}%
\providecommand \@sanitize@url [0]{\catcode `\\12\catcode `\$12\catcode
  `\&12\catcode `\#12\catcode `\^12\catcode `\_12\catcode `\%12\relax}%
\providecommand \@@startlink[1]{}%
\providecommand \@@endlink[0]{}%
\providecommand \url  [0]{\begingroup\@sanitize@url \@url }%
\providecommand \@url [1]{\endgroup\@href {#1}{\urlprefix }}%
\providecommand \urlprefix  [0]{URL }%
\providecommand \Eprint [0]{\href }%
\providecommand \doibase [0]{https://doi.org/}%
\providecommand \selectlanguage [0]{\@gobble}%
\providecommand \bibinfo  [0]{\@secondoftwo}%
\providecommand \bibfield  [0]{\@secondoftwo}%
\providecommand \translation [1]{[#1]}%
\providecommand \BibitemOpen [0]{}%
\providecommand \bibitemStop [0]{}%
\providecommand \bibitemNoStop [0]{.\EOS\space}%
\providecommand \EOS [0]{\spacefactor3000\relax}%
\providecommand \BibitemShut  [1]{\csname bibitem#1\endcsname}%
\let\auto@bib@innerbib\@empty
\bibitem [{\citenamefont {Haldane}(1926)}]{haldane1926being}%
  \BibitemOpen
  \bibfield  {author} {\bibinfo {author} {\bibfnamefont {J.~B.}\ \bibnamefont
  {Haldane}},\ }\bibfield  {title} {\bibinfo {title} {On being the right
  size},\ }\href@noop {} {\bibfield  {journal} {\bibinfo  {journal} {Harper’s
  magazine}\ }\textbf {\bibinfo {volume} {152}},\ \bibinfo {pages} {424}
  (\bibinfo {year} {1926})}\BibitemShut {NoStop}%
\bibitem [{\citenamefont {Ginzberg}\ \emph {et~al.}(2015)\citenamefont
  {Ginzberg}, \citenamefont {Kafri},\ and\ \citenamefont
  {Kirschner}}]{ginzberg2015being}%
  \BibitemOpen
  \bibfield  {author} {\bibinfo {author} {\bibfnamefont {M.~B.}\ \bibnamefont
  {Ginzberg}}, \bibinfo {author} {\bibfnamefont {R.}~\bibnamefont {Kafri}},\
  and\ \bibinfo {author} {\bibfnamefont {M.}~\bibnamefont {Kirschner}},\
  }\bibfield  {title} {\bibinfo {title} {On being the right (cell) size},\
  }\href@noop {} {\bibfield  {journal} {\bibinfo  {journal} {Science}\ }\textbf
  {\bibinfo {volume} {348}},\ \bibinfo {pages} {1245075} (\bibinfo {year}
  {2015})}\BibitemShut {NoStop}%
\bibitem [{\citenamefont {Banerjee}\ and\ \citenamefont
  {Banerjee}(2025)}]{banerjee2025design}%
  \BibitemOpen
  \bibfield  {author} {\bibinfo {author} {\bibfnamefont {D.~S.}\ \bibnamefont
  {Banerjee}}\ and\ \bibinfo {author} {\bibfnamefont {S.}~\bibnamefont
  {Banerjee}},\ }\bibfield  {title} {\bibinfo {title} {Design principles and
  feedback mechanisms in organelle size control},\ }\href@noop {} {\bibfield
  {journal} {\bibinfo  {journal} {Current Opinion in Cell Biology}\ }\textbf
  {\bibinfo {volume} {95}},\ \bibinfo {pages} {102533} (\bibinfo {year}
  {2025})}\BibitemShut {NoStop}%
\bibitem [{\citenamefont {Boeynaems}\ \emph {et~al.}(2018)\citenamefont
  {Boeynaems}, \citenamefont {Alberti}, \citenamefont {Fawzi}, \citenamefont
  {Mittag}, \citenamefont {Polymenidou}, \citenamefont {Rousseau},
  \citenamefont {Schymkowitz}, \citenamefont {Shorter}, \citenamefont
  {Wolozin}, \citenamefont {Van Den~Bosch} \emph
  {et~al.}}]{boeynaems2018protein}%
  \BibitemOpen
  \bibfield  {author} {\bibinfo {author} {\bibfnamefont {S.}~\bibnamefont
  {Boeynaems}}, \bibinfo {author} {\bibfnamefont {S.}~\bibnamefont {Alberti}},
  \bibinfo {author} {\bibfnamefont {N.~L.}\ \bibnamefont {Fawzi}}, \bibinfo
  {author} {\bibfnamefont {T.}~\bibnamefont {Mittag}}, \bibinfo {author}
  {\bibfnamefont {M.}~\bibnamefont {Polymenidou}}, \bibinfo {author}
  {\bibfnamefont {F.}~\bibnamefont {Rousseau}}, \bibinfo {author}
  {\bibfnamefont {J.}~\bibnamefont {Schymkowitz}}, \bibinfo {author}
  {\bibfnamefont {J.}~\bibnamefont {Shorter}}, \bibinfo {author} {\bibfnamefont
  {B.}~\bibnamefont {Wolozin}}, \bibinfo {author} {\bibfnamefont
  {L.}~\bibnamefont {Van Den~Bosch}}, \emph {et~al.},\ }\bibfield  {title}
  {\bibinfo {title} {Protein phase separation: a new phase in cell biology},\
  }\href@noop {} {\bibfield  {journal} {\bibinfo  {journal} {Trends in cell
  biology}\ }\textbf {\bibinfo {volume} {28}},\ \bibinfo {pages} {420}
  (\bibinfo {year} {2018})}\BibitemShut {NoStop}%
\bibitem [{\citenamefont {Mitrea}\ and\ \citenamefont
  {Kriwacki}(2016)}]{mitrea2016phase}%
  \BibitemOpen
  \bibfield  {author} {\bibinfo {author} {\bibfnamefont {D.~M.}\ \bibnamefont
  {Mitrea}}\ and\ \bibinfo {author} {\bibfnamefont {R.~W.}\ \bibnamefont
  {Kriwacki}},\ }\bibfield  {title} {\bibinfo {title} {Phase separation in
  biology; functional organization of a higher order},\ }\href@noop {}
  {\bibfield  {journal} {\bibinfo  {journal} {Cell Communication and
  Signaling}\ }\textbf {\bibinfo {volume} {14}},\ \bibinfo {pages} {1}
  (\bibinfo {year} {2016})}\BibitemShut {NoStop}%
\bibitem [{\citenamefont {Zwicker}\ \emph {et~al.}(2025)\citenamefont
  {Zwicker}, \citenamefont {Paulin},\ and\ \citenamefont {ter
  Burg}}]{zwicker2025physics}%
  \BibitemOpen
  \bibfield  {author} {\bibinfo {author} {\bibfnamefont {D.}~\bibnamefont
  {Zwicker}}, \bibinfo {author} {\bibfnamefont {O.~W.}\ \bibnamefont
  {Paulin}},\ and\ \bibinfo {author} {\bibfnamefont {C.}~\bibnamefont {ter
  Burg}},\ }\bibfield  {title} {\bibinfo {title} {Physics of droplet regulation
  in biological cells},\ }\href@noop {} {\bibfield  {journal} {\bibinfo
  {journal} {arXiv preprint arXiv:2501.13639}\ } (\bibinfo {year}
  {2025})}\BibitemShut {NoStop}%
\bibitem [{\citenamefont {Weber}\ \emph {et~al.}(2019)\citenamefont {Weber},
  \citenamefont {Zwicker}, \citenamefont {J{\"u}licher},\ and\ \citenamefont
  {Lee}}]{weber2019physics}%
  \BibitemOpen
  \bibfield  {author} {\bibinfo {author} {\bibfnamefont {C.~A.}\ \bibnamefont
  {Weber}}, \bibinfo {author} {\bibfnamefont {D.}~\bibnamefont {Zwicker}},
  \bibinfo {author} {\bibfnamefont {F.}~\bibnamefont {J{\"u}licher}},\ and\
  \bibinfo {author} {\bibfnamefont {C.~F.}\ \bibnamefont {Lee}},\ }\bibfield
  {title} {\bibinfo {title} {Physics of active emulsions},\ }\href@noop {}
  {\bibfield  {journal} {\bibinfo  {journal} {Reports on Progress in Physics}\
  }\textbf {\bibinfo {volume} {82}},\ \bibinfo {pages} {064601} (\bibinfo
  {year} {2019})}\BibitemShut {NoStop}%
\bibitem [{\citenamefont {Turing}(1952)}]{turing1952chemical}%
  \BibitemOpen
  \bibfield  {author} {\bibinfo {author} {\bibfnamefont {A.~M.}\ \bibnamefont
  {Turing}},\ }\bibfield  {title} {\bibinfo {title} {The chemical basis of
  morphogenesis},\ }\href@noop {} {\bibfield  {journal} {\bibinfo  {journal}
  {Philosophical Transactions of the Royal Society of London. Series B,
  Biological Sciences}\ }\textbf {\bibinfo {volume} {237}},\ \bibinfo {pages}
  {37} (\bibinfo {year} {1952})}\BibitemShut {NoStop}%
\bibitem [{\citenamefont {Kondo}\ and\ \citenamefont
  {Miura}(2010)}]{kondo2010reaction}%
  \BibitemOpen
  \bibfield  {author} {\bibinfo {author} {\bibfnamefont {S.}~\bibnamefont
  {Kondo}}\ and\ \bibinfo {author} {\bibfnamefont {T.}~\bibnamefont {Miura}},\
  }\bibfield  {title} {\bibinfo {title} {Reaction-diffusion model as a
  framework for understanding biological pattern formation},\ }\href@noop {}
  {\bibfield  {journal} {\bibinfo  {journal} {science}\ }\textbf {\bibinfo
  {volume} {329}},\ \bibinfo {pages} {1616} (\bibinfo {year}
  {2010})}\BibitemShut {NoStop}%
\bibitem [{\citenamefont {Howard}\ \emph {et~al.}(2011)\citenamefont {Howard},
  \citenamefont {Grill},\ and\ \citenamefont {Bois}}]{howard2011turing}%
  \BibitemOpen
  \bibfield  {author} {\bibinfo {author} {\bibfnamefont {J.}~\bibnamefont
  {Howard}}, \bibinfo {author} {\bibfnamefont {S.~W.}\ \bibnamefont {Grill}},\
  and\ \bibinfo {author} {\bibfnamefont {J.~S.}\ \bibnamefont {Bois}},\
  }\bibfield  {title} {\bibinfo {title} {Turing's next steps: the
  mechanochemical basis of morphogenesis},\ }\href@noop {} {\bibfield
  {journal} {\bibinfo  {journal} {Nature Reviews Molecular Cell Biology}\
  }\textbf {\bibinfo {volume} {12}},\ \bibinfo {pages} {392} (\bibinfo {year}
  {2011})}\BibitemShut {NoStop}%
\bibitem [{\citenamefont {Ge{\ss}ele}\ \emph {et~al.}(2020)\citenamefont
  {Ge{\ss}ele}, \citenamefont {Halatek}, \citenamefont {W{\"u}rthner},\ and\
  \citenamefont {Frey}}]{gessele2020geometric}%
  \BibitemOpen
  \bibfield  {author} {\bibinfo {author} {\bibfnamefont {R.}~\bibnamefont
  {Ge{\ss}ele}}, \bibinfo {author} {\bibfnamefont {J.}~\bibnamefont {Halatek}},
  \bibinfo {author} {\bibfnamefont {L.}~\bibnamefont {W{\"u}rthner}},\ and\
  \bibinfo {author} {\bibfnamefont {E.}~\bibnamefont {Frey}},\ }\bibfield
  {title} {\bibinfo {title} {Geometric cues stabilise long-axis polarisation of
  par protein patterns in c. elegans},\ }\href@noop {} {\bibfield  {journal}
  {\bibinfo  {journal} {Nature communications}\ }\textbf {\bibinfo {volume}
  {11}},\ \bibinfo {pages} {539} (\bibinfo {year} {2020})}\BibitemShut
  {NoStop}%
\bibitem [{\citenamefont {Bergmann}\ \emph {et~al.}(2018)\citenamefont
  {Bergmann}, \citenamefont {Rapp},\ and\ \citenamefont
  {Zimmermann}}]{bergmann2018active}%
  \BibitemOpen
  \bibfield  {author} {\bibinfo {author} {\bibfnamefont {F.}~\bibnamefont
  {Bergmann}}, \bibinfo {author} {\bibfnamefont {L.}~\bibnamefont {Rapp}},\
  and\ \bibinfo {author} {\bibfnamefont {W.}~\bibnamefont {Zimmermann}},\
  }\bibfield  {title} {\bibinfo {title} {Active phase separation: A universal
  approach},\ }\href@noop {} {\bibfield  {journal} {\bibinfo  {journal}
  {Physical Review E}\ }\textbf {\bibinfo {volume} {98}},\ \bibinfo {pages}
  {020603} (\bibinfo {year} {2018})}\BibitemShut {NoStop}%
\bibitem [{\citenamefont {Robinson}\ \emph {et~al.}(2025)\citenamefont
  {Robinson}, \citenamefont {Machon},\ and\ \citenamefont
  {Speck}}]{robinson2025universal}%
  \BibitemOpen
  \bibfield  {author} {\bibinfo {author} {\bibfnamefont {J.~F.}\ \bibnamefont
  {Robinson}}, \bibinfo {author} {\bibfnamefont {T.}~\bibnamefont {Machon}},\
  and\ \bibinfo {author} {\bibfnamefont {T.}~\bibnamefont {Speck}},\ }\bibfield
   {title} {\bibinfo {title} {Universal limiting behavior of reaction-diffusion
  systems with conservation laws},\ }\href@noop {} {\bibfield  {journal}
  {\bibinfo  {journal} {Physical Review E}\ }\textbf {\bibinfo {volume}
  {111}},\ \bibinfo {pages} {065417} (\bibinfo {year} {2025})}\BibitemShut
  {NoStop}%
\bibitem [{\citenamefont {Cahn}\ and\ \citenamefont
  {Hilliard}(1958)}]{cahn1958free}%
  \BibitemOpen
  \bibfield  {author} {\bibinfo {author} {\bibfnamefont {J.~W.}\ \bibnamefont
  {Cahn}}\ and\ \bibinfo {author} {\bibfnamefont {J.~E.}\ \bibnamefont
  {Hilliard}},\ }\bibfield  {title} {\bibinfo {title} {Free energy of a
  nonuniform system. i. interfacial free energy},\ }\href@noop {} {\bibfield
  {journal} {\bibinfo  {journal} {The Journal of chemical physics}\ }\textbf
  {\bibinfo {volume} {28}},\ \bibinfo {pages} {258} (\bibinfo {year}
  {1958})}\BibitemShut {NoStop}%
\bibitem [{\citenamefont {Cahn}(1965)}]{cahn1965phase}%
  \BibitemOpen
  \bibfield  {author} {\bibinfo {author} {\bibfnamefont {J.~W.}\ \bibnamefont
  {Cahn}},\ }\bibfield  {title} {\bibinfo {title} {Phase separation by spinodal
  decomposition in isotropic systems},\ }\href@noop {} {\bibfield  {journal}
  {\bibinfo  {journal} {The Journal of chemical physics}\ }\textbf {\bibinfo
  {volume} {42}},\ \bibinfo {pages} {93} (\bibinfo {year} {1965})}\BibitemShut
  {NoStop}%
\bibitem [{\citenamefont {Novick-Cohen}(1985)}]{novick1985nonlinear}%
  \BibitemOpen
  \bibfield  {author} {\bibinfo {author} {\bibfnamefont {A.}~\bibnamefont
  {Novick-Cohen}},\ }\bibfield  {title} {\bibinfo {title} {The nonlinear
  cahn-hilliard equation: transition from spinodal decomposition to nucleation
  behavior},\ }\href@noop {} {\bibfield  {journal} {\bibinfo  {journal}
  {Journal of statistical physics}\ }\textbf {\bibinfo {volume} {38}},\
  \bibinfo {pages} {707} (\bibinfo {year} {1985})}\BibitemShut {NoStop}%
\bibitem [{\citenamefont {Bray}(2003)}]{bray2003coarsening}%
  \BibitemOpen
  \bibfield  {author} {\bibinfo {author} {\bibfnamefont {A.}~\bibnamefont
  {Bray}},\ }\bibfield  {title} {\bibinfo {title} {Coarsening dynamics of
  phase-separating systems},\ }\href@noop {} {\bibfield  {journal} {\bibinfo
  {journal} {Philosophical Transactions of the Royal Society of London. Series
  A: Mathematical, Physical and Engineering Sciences}\ }\textbf {\bibinfo
  {volume} {361}},\ \bibinfo {pages} {781} (\bibinfo {year}
  {2003})}\BibitemShut {NoStop}%
\bibitem [{\citenamefont {Ostwald}(1903)}]{ostwald1903lehrbuch}%
  \BibitemOpen
  \bibfield  {author} {\bibinfo {author} {\bibfnamefont {W.}~\bibnamefont
  {Ostwald}},\ }\href@noop {} {\emph {\bibinfo {title} {Lehrbuch der
  allgemeinen Chemie}}},\ Vol.~\bibinfo {volume} {1}\ (\bibinfo  {publisher}
  {W. Engelmann},\ \bibinfo {year} {1903})\BibitemShut {NoStop}%
\bibitem [{\citenamefont {Brauns}\ \emph {et~al.}(2021)\citenamefont {Brauns},
  \citenamefont {Weyer}, \citenamefont {Halatek}, \citenamefont {Yoon},\ and\
  \citenamefont {Frey}}]{brauns2021wavelength}%
  \BibitemOpen
  \bibfield  {author} {\bibinfo {author} {\bibfnamefont {F.}~\bibnamefont
  {Brauns}}, \bibinfo {author} {\bibfnamefont {H.}~\bibnamefont {Weyer}},
  \bibinfo {author} {\bibfnamefont {J.}~\bibnamefont {Halatek}}, \bibinfo
  {author} {\bibfnamefont {J.}~\bibnamefont {Yoon}},\ and\ \bibinfo {author}
  {\bibfnamefont {E.}~\bibnamefont {Frey}},\ }\bibfield  {title} {\bibinfo
  {title} {Wavelength selection by interrupted coarsening in reaction-diffusion
  systems},\ }\href@noop {} {\bibfield  {journal} {\bibinfo  {journal}
  {Physical Review Letters}\ }\textbf {\bibinfo {volume} {126}},\ \bibinfo
  {pages} {104101} (\bibinfo {year} {2021})}\BibitemShut {NoStop}%
\bibitem [{\citenamefont {Weyer}\ \emph {et~al.}(2023)\citenamefont {Weyer},
  \citenamefont {Brauns},\ and\ \citenamefont {Frey}}]{weyer2023coarsening}%
  \BibitemOpen
  \bibfield  {author} {\bibinfo {author} {\bibfnamefont {H.}~\bibnamefont
  {Weyer}}, \bibinfo {author} {\bibfnamefont {F.}~\bibnamefont {Brauns}},\ and\
  \bibinfo {author} {\bibfnamefont {E.}~\bibnamefont {Frey}},\ }\bibfield
  {title} {\bibinfo {title} {Coarsening and wavelength selection far from
  equilibrium: a unifying framework based on singular perturbation theory},\
  }\href@noop {} {\bibfield  {journal} {\bibinfo  {journal} {Physical Review
  E}\ }\textbf {\bibinfo {volume} {108}},\ \bibinfo {pages} {064202} (\bibinfo
  {year} {2023})}\BibitemShut {NoStop}%
\bibitem [{\citenamefont {Gonnella}\ \emph {et~al.}(2015)\citenamefont
  {Gonnella}, \citenamefont {Marenduzzo}, \citenamefont {Suma},\ and\
  \citenamefont {Tiribocchi}}]{gonnella2015motility}%
  \BibitemOpen
  \bibfield  {author} {\bibinfo {author} {\bibfnamefont {G.}~\bibnamefont
  {Gonnella}}, \bibinfo {author} {\bibfnamefont {D.}~\bibnamefont
  {Marenduzzo}}, \bibinfo {author} {\bibfnamefont {A.}~\bibnamefont {Suma}},\
  and\ \bibinfo {author} {\bibfnamefont {A.}~\bibnamefont {Tiribocchi}},\
  }\bibfield  {title} {\bibinfo {title} {Motility-induced phase separation and
  coarsening in active matter},\ }\href@noop {} {\bibfield  {journal} {\bibinfo
   {journal} {Comptes Rendus. Physique}\ }\textbf {\bibinfo {volume} {16}},\
  \bibinfo {pages} {316} (\bibinfo {year} {2015})}\BibitemShut {NoStop}%
\bibitem [{\citenamefont {Zwicker}\ \emph {et~al.}(2015)\citenamefont
  {Zwicker}, \citenamefont {Hyman},\ and\ \citenamefont
  {J{\"u}licher}}]{zwicker2015suppression}%
  \BibitemOpen
  \bibfield  {author} {\bibinfo {author} {\bibfnamefont {D.}~\bibnamefont
  {Zwicker}}, \bibinfo {author} {\bibfnamefont {A.~A.}\ \bibnamefont {Hyman}},\
  and\ \bibinfo {author} {\bibfnamefont {F.}~\bibnamefont {J{\"u}licher}},\
  }\bibfield  {title} {\bibinfo {title} {Suppression of ostwald ripening in
  active emulsions},\ }\href@noop {} {\bibfield  {journal} {\bibinfo  {journal}
  {Physical Review E}\ }\textbf {\bibinfo {volume} {92}},\ \bibinfo {pages}
  {012317} (\bibinfo {year} {2015})}\BibitemShut {NoStop}%
\bibitem [{\citenamefont {Zwicker}(2022)}]{zwicker2022intertwined}%
  \BibitemOpen
  \bibfield  {author} {\bibinfo {author} {\bibfnamefont {D.}~\bibnamefont
  {Zwicker}},\ }\bibfield  {title} {\bibinfo {title} {The intertwined physics
  of active chemical reactions and phase separation},\ }\href@noop {}
  {\bibfield  {journal} {\bibinfo  {journal} {Current Opinion in Colloid \&
  Interface Science}\ }\textbf {\bibinfo {volume} {61}},\ \bibinfo {pages}
  {101606} (\bibinfo {year} {2022})}\BibitemShut {NoStop}%
\bibitem [{\citenamefont {Garcke}(2003)}]{garcke2003cahn}%
  \BibitemOpen
  \bibfield  {author} {\bibinfo {author} {\bibfnamefont {H.}~\bibnamefont
  {Garcke}},\ }\bibfield  {title} {\bibinfo {title} {On cahn—hilliard systems
  with elasticity},\ }\href@noop {} {\bibfield  {journal} {\bibinfo  {journal}
  {Proceedings of the Royal Society of Edinburgh Section A: Mathematics}\
  }\textbf {\bibinfo {volume} {133}},\ \bibinfo {pages} {307} (\bibinfo {year}
  {2003})}\BibitemShut {NoStop}%
\bibitem [{\citenamefont {Padhan}\ \emph {et~al.}(2024)\citenamefont {Padhan},
  \citenamefont {Kiran},\ and\ \citenamefont {Pandit}}]{padhan2024novel}%
  \BibitemOpen
  \bibfield  {author} {\bibinfo {author} {\bibfnamefont {N.~B.}\ \bibnamefont
  {Padhan}}, \bibinfo {author} {\bibfnamefont {K.~V.}\ \bibnamefont {Kiran}},\
  and\ \bibinfo {author} {\bibfnamefont {R.}~\bibnamefont {Pandit}},\
  }\bibfield  {title} {\bibinfo {title} {Novel turbulence and coarsening arrest
  in active-scalar fluids},\ }\href@noop {} {\bibfield  {journal} {\bibinfo
  {journal} {Soft Matter}\ }\textbf {\bibinfo {volume} {20}},\ \bibinfo {pages}
  {3620} (\bibinfo {year} {2024})}\BibitemShut {NoStop}%
\bibitem [{\citenamefont {Frohoff-H{\"u}lsmann}\ \emph
  {et~al.}(2021)\citenamefont {Frohoff-H{\"u}lsmann}, \citenamefont {Wrembel},\
  and\ \citenamefont {Thiele}}]{frohoff2021suppression}%
  \BibitemOpen
  \bibfield  {author} {\bibinfo {author} {\bibfnamefont {T.}~\bibnamefont
  {Frohoff-H{\"u}lsmann}}, \bibinfo {author} {\bibfnamefont {J.}~\bibnamefont
  {Wrembel}},\ and\ \bibinfo {author} {\bibfnamefont {U.}~\bibnamefont
  {Thiele}},\ }\bibfield  {title} {\bibinfo {title} {Suppression of coarsening
  and emergence of oscillatory behavior in a cahn-hilliard model with
  nonvariational coupling},\ }\href@noop {} {\bibfield  {journal} {\bibinfo
  {journal} {Physical Review E}\ }\textbf {\bibinfo {volume} {103}},\ \bibinfo
  {pages} {042602} (\bibinfo {year} {2021})}\BibitemShut {NoStop}%
\bibitem [{\citenamefont {Frohoff-H{\"u}lsmann}\ \emph
  {et~al.}(2023)\citenamefont {Frohoff-H{\"u}lsmann}, \citenamefont {Thiele},\
  and\ \citenamefont {Pismen}}]{frohoff2023non}%
  \BibitemOpen
  \bibfield  {author} {\bibinfo {author} {\bibfnamefont {T.}~\bibnamefont
  {Frohoff-H{\"u}lsmann}}, \bibinfo {author} {\bibfnamefont {U.}~\bibnamefont
  {Thiele}},\ and\ \bibinfo {author} {\bibfnamefont {L.~M.}\ \bibnamefont
  {Pismen}},\ }\bibfield  {title} {\bibinfo {title} {Non-reciprocity induces
  resonances in a two-field cahn--hilliard model},\ }\href@noop {} {\bibfield
  {journal} {\bibinfo  {journal} {Philosophical Transactions of the Royal
  Society A}\ }\textbf {\bibinfo {volume} {381}},\ \bibinfo {pages} {20220087}
  (\bibinfo {year} {2023})}\BibitemShut {NoStop}%
\bibitem [{\citenamefont {Mao}\ \emph {et~al.}(2014)\citenamefont {Mao},
  \citenamefont {Lou}, \citenamefont {Lou}, \citenamefont {Wang},\ and\
  \citenamefont {Jin}}]{mao2014behaviour}%
  \BibitemOpen
  \bibfield  {author} {\bibinfo {author} {\bibfnamefont {L.}~\bibnamefont
  {Mao}}, \bibinfo {author} {\bibfnamefont {H.}~\bibnamefont {Lou}}, \bibinfo
  {author} {\bibfnamefont {Y.}~\bibnamefont {Lou}}, \bibinfo {author}
  {\bibfnamefont {N.}~\bibnamefont {Wang}},\ and\ \bibinfo {author}
  {\bibfnamefont {F.}~\bibnamefont {Jin}},\ }\bibfield  {title} {\bibinfo
  {title} {Behaviour of cytoplasmic organelles and cytoskeleton during oocyte
  maturation},\ }\href@noop {} {\bibfield  {journal} {\bibinfo  {journal}
  {Reproductive biomedicine online}\ }\textbf {\bibinfo {volume} {28}},\
  \bibinfo {pages} {284} (\bibinfo {year} {2014})}\BibitemShut {NoStop}%
\bibitem [{\citenamefont {Chen}\ and\ \citenamefont
  {Levy}(2023)}]{chen2023regulation}%
  \BibitemOpen
  \bibfield  {author} {\bibinfo {author} {\bibfnamefont {P.}~\bibnamefont
  {Chen}}\ and\ \bibinfo {author} {\bibfnamefont {D.~L.}\ \bibnamefont
  {Levy}},\ }\bibfield  {title} {\bibinfo {title} {Regulation of organelle size
  and organization during development},\ }in\ \href@noop {} {\emph {\bibinfo
  {booktitle} {Seminars in cell \& developmental biology}}},\ Vol.\ \bibinfo
  {volume} {133}\ (\bibinfo {organization} {Elsevier},\ \bibinfo {year}
  {2023})\ pp.\ \bibinfo {pages} {53--64}\BibitemShut {NoStop}%
\bibitem [{\citenamefont {Alberts}\ \emph {et~al.}(2008)\citenamefont
  {Alberts}, \citenamefont {Wilson},\ and\ \citenamefont
  {Hunt}}]{alberts2008molecular}%
  \BibitemOpen
  \bibfield  {author} {\bibinfo {author} {\bibfnamefont {B.}~\bibnamefont
  {Alberts}}, \bibinfo {author} {\bibfnamefont {J.}~\bibnamefont {Wilson}},\
  and\ \bibinfo {author} {\bibfnamefont {T.}~\bibnamefont {Hunt}},\ }\href@noop
  {} {\bibinfo {title} {Molecular biology of the cell (garland science, new
  york)}} (\bibinfo {year} {2008})\BibitemShut {NoStop}%
\bibitem [{\citenamefont {Needleman}\ \emph {et~al.}(2010)\citenamefont
  {Needleman}, \citenamefont {Groen}, \citenamefont {Ohi}, \citenamefont
  {Maresca}, \citenamefont {Mirny},\ and\ \citenamefont
  {Mitchison}}]{needleman2010fast}%
  \BibitemOpen
  \bibfield  {author} {\bibinfo {author} {\bibfnamefont {D.~J.}\ \bibnamefont
  {Needleman}}, \bibinfo {author} {\bibfnamefont {A.}~\bibnamefont {Groen}},
  \bibinfo {author} {\bibfnamefont {R.}~\bibnamefont {Ohi}}, \bibinfo {author}
  {\bibfnamefont {T.}~\bibnamefont {Maresca}}, \bibinfo {author} {\bibfnamefont
  {L.}~\bibnamefont {Mirny}},\ and\ \bibinfo {author} {\bibfnamefont
  {T.}~\bibnamefont {Mitchison}},\ }\bibfield  {title} {\bibinfo {title} {Fast
  microtubule dynamics in meiotic spindles measured by single molecule imaging:
  evidence that the spindle environment does not stabilize microtubules},\
  }\href@noop {} {\bibfield  {journal} {\bibinfo  {journal} {Molecular biology
  of the cell}\ }\textbf {\bibinfo {volume} {21}},\ \bibinfo {pages} {323}
  (\bibinfo {year} {2010})}\BibitemShut {NoStop}%
\bibitem [{\citenamefont {Fritzsche}\ \emph {et~al.}(2013)\citenamefont
  {Fritzsche}, \citenamefont {Lewalle}, \citenamefont {Duke}, \citenamefont
  {Kruse},\ and\ \citenamefont {Charras}}]{fritzsche2013analysis}%
  \BibitemOpen
  \bibfield  {author} {\bibinfo {author} {\bibfnamefont {M.}~\bibnamefont
  {Fritzsche}}, \bibinfo {author} {\bibfnamefont {A.}~\bibnamefont {Lewalle}},
  \bibinfo {author} {\bibfnamefont {T.}~\bibnamefont {Duke}}, \bibinfo {author}
  {\bibfnamefont {K.}~\bibnamefont {Kruse}},\ and\ \bibinfo {author}
  {\bibfnamefont {G.}~\bibnamefont {Charras}},\ }\bibfield  {title} {\bibinfo
  {title} {Analysis of turnover dynamics of the submembranous actin cortex},\
  }\href@noop {} {\bibfield  {journal} {\bibinfo  {journal} {Molecular biology
  of the cell}\ }\textbf {\bibinfo {volume} {24}},\ \bibinfo {pages} {757}
  (\bibinfo {year} {2013})}\BibitemShut {NoStop}%
\bibitem [{\citenamefont {Petry}\ \emph {et~al.}(2013)\citenamefont {Petry},
  \citenamefont {Groen}, \citenamefont {Ishihara}, \citenamefont {Mitchison},\
  and\ \citenamefont {Vale}}]{petry2013branching}%
  \BibitemOpen
  \bibfield  {author} {\bibinfo {author} {\bibfnamefont {S.}~\bibnamefont
  {Petry}}, \bibinfo {author} {\bibfnamefont {A.~C.}\ \bibnamefont {Groen}},
  \bibinfo {author} {\bibfnamefont {K.}~\bibnamefont {Ishihara}}, \bibinfo
  {author} {\bibfnamefont {T.~J.}\ \bibnamefont {Mitchison}},\ and\ \bibinfo
  {author} {\bibfnamefont {R.~D.}\ \bibnamefont {Vale}},\ }\bibfield  {title}
  {\bibinfo {title} {Branching microtubule nucleation in xenopus egg extracts
  mediated by augmin and tpx2},\ }\href@noop {} {\bibfield  {journal} {\bibinfo
   {journal} {Cell}\ }\textbf {\bibinfo {volume} {152}},\ \bibinfo {pages}
  {768} (\bibinfo {year} {2013})}\BibitemShut {NoStop}%
\bibitem [{\citenamefont {Kaye}\ \emph {et~al.}(2018)\citenamefont {Kaye},
  \citenamefont {Stiehl}, \citenamefont {Foster}, \citenamefont {Shelley},
  \citenamefont {Needleman},\ and\ \citenamefont
  {F{\"u}rthauer}}]{kaye2018measuring}%
  \BibitemOpen
  \bibfield  {author} {\bibinfo {author} {\bibfnamefont {B.}~\bibnamefont
  {Kaye}}, \bibinfo {author} {\bibfnamefont {O.}~\bibnamefont {Stiehl}},
  \bibinfo {author} {\bibfnamefont {P.~J.}\ \bibnamefont {Foster}}, \bibinfo
  {author} {\bibfnamefont {M.~J.}\ \bibnamefont {Shelley}}, \bibinfo {author}
  {\bibfnamefont {D.~J.}\ \bibnamefont {Needleman}},\ and\ \bibinfo {author}
  {\bibfnamefont {S.}~\bibnamefont {F{\"u}rthauer}},\ }\bibfield  {title}
  {\bibinfo {title} {Measuring and modeling polymer concentration profiles near
  spindle boundaries argues that spindle microtubules regulate their own
  nucleation},\ }\href@noop {} {\bibfield  {journal} {\bibinfo  {journal} {New
  Journal of Physics}\ }\textbf {\bibinfo {volume} {20}},\ \bibinfo {pages}
  {055012} (\bibinfo {year} {2018})}\BibitemShut {NoStop}%
\bibitem [{\citenamefont {F{\"u}rthauer}\ \emph {et~al.}(2019)\citenamefont
  {F{\"u}rthauer}, \citenamefont {Lemma}, \citenamefont {Foster}, \citenamefont
  {Ems-McClung}, \citenamefont {Yu}, \citenamefont {Walczak}, \citenamefont
  {Dogic}, \citenamefont {Needleman},\ and\ \citenamefont
  {Shelley}}]{furthauer2019self}%
  \BibitemOpen
  \bibfield  {author} {\bibinfo {author} {\bibfnamefont {S.}~\bibnamefont
  {F{\"u}rthauer}}, \bibinfo {author} {\bibfnamefont {B.}~\bibnamefont
  {Lemma}}, \bibinfo {author} {\bibfnamefont {P.~J.}\ \bibnamefont {Foster}},
  \bibinfo {author} {\bibfnamefont {S.~C.}\ \bibnamefont {Ems-McClung}},
  \bibinfo {author} {\bibfnamefont {C.-H.}\ \bibnamefont {Yu}}, \bibinfo
  {author} {\bibfnamefont {C.~E.}\ \bibnamefont {Walczak}}, \bibinfo {author}
  {\bibfnamefont {Z.}~\bibnamefont {Dogic}}, \bibinfo {author} {\bibfnamefont
  {D.~J.}\ \bibnamefont {Needleman}},\ and\ \bibinfo {author} {\bibfnamefont
  {M.~J.}\ \bibnamefont {Shelley}},\ }\bibfield  {title} {\bibinfo {title}
  {Self-straining of actively crosslinked microtubule networks},\ }\href@noop
  {} {\bibfield  {journal} {\bibinfo  {journal} {Nature physics}\ }\textbf
  {\bibinfo {volume} {15}},\ \bibinfo {pages} {1295} (\bibinfo {year}
  {2019})}\BibitemShut {NoStop}%
\bibitem [{\citenamefont {F{\"u}rthauer}\ \emph {et~al.}(2021)\citenamefont
  {F{\"u}rthauer}, \citenamefont {Needleman},\ and\ \citenamefont
  {Shelley}}]{furthauer2021design}%
  \BibitemOpen
  \bibfield  {author} {\bibinfo {author} {\bibfnamefont {S.}~\bibnamefont
  {F{\"u}rthauer}}, \bibinfo {author} {\bibfnamefont {D.~J.}\ \bibnamefont
  {Needleman}},\ and\ \bibinfo {author} {\bibfnamefont {M.~J.}\ \bibnamefont
  {Shelley}},\ }\bibfield  {title} {\bibinfo {title} {A design framework for
  actively crosslinked filament networks},\ }\href@noop {} {\bibfield
  {journal} {\bibinfo  {journal} {New Journal of Physics}\ }\textbf {\bibinfo
  {volume} {23}},\ \bibinfo {pages} {013012} (\bibinfo {year}
  {2021})}\BibitemShut {NoStop}%
\bibitem [{\citenamefont {Zampetaki}(2025)}]{zampetaki2025}%
  \BibitemOpen
  \bibfield  {author} {\bibinfo {author} {\bibfnamefont {A.}~\bibnamefont
  {Zampetaki}},\ }\bibfield  {title} {\bibinfo {title} {To be published}}
  (\bibinfo {year} {2025})\BibitemShut {NoStop}%
\bibitem [{\citenamefont {Ginzburg}\ and\ \citenamefont
  {Landau}(1950)}]{ginzburg1950theory}%
  \BibitemOpen
  \bibfield  {author} {\bibinfo {author} {\bibfnamefont {V.}~\bibnamefont
  {Ginzburg}}\ and\ \bibinfo {author} {\bibfnamefont {L.}~\bibnamefont
  {Landau}},\ }\bibfield  {title} {\bibinfo {title} {Theory of
  superconductivity},\ }\href@noop {} {\bibfield  {journal} {\bibinfo
  {journal} {Zh. Eksp. Teor. Fiz.;(USSR)}\ }\textbf {\bibinfo {volume} {20}}
  (\bibinfo {year} {1950})}\BibitemShut {NoStop}%
\bibitem [{\citenamefont {Burns}\ \emph {et~al.}(2020)\citenamefont {Burns},
  \citenamefont {Vasil}, \citenamefont {Oishi}, \citenamefont {Lecoanet},\ and\
  \citenamefont {Brown}}]{burns2020dedalus}%
  \BibitemOpen
  \bibfield  {author} {\bibinfo {author} {\bibfnamefont {K.~J.}\ \bibnamefont
  {Burns}}, \bibinfo {author} {\bibfnamefont {G.~M.}\ \bibnamefont {Vasil}},
  \bibinfo {author} {\bibfnamefont {J.~S.}\ \bibnamefont {Oishi}}, \bibinfo
  {author} {\bibfnamefont {D.}~\bibnamefont {Lecoanet}},\ and\ \bibinfo
  {author} {\bibfnamefont {B.~P.}\ \bibnamefont {Brown}},\ }\bibfield  {title}
  {\bibinfo {title} {Dedalus: A flexible framework for numerical simulations
  with spectral methods},\ }\href@noop {} {\bibfield  {journal} {\bibinfo
  {journal} {Physical Review Research}\ }\textbf {\bibinfo {volume} {2}},\
  \bibinfo {pages} {023068} (\bibinfo {year} {2020})}\BibitemShut {NoStop}%
\bibitem [{\citenamefont {Bodini-Lefranc}\ \emph {et~al.}(2025)\citenamefont
  {Bodini-Lefranc}, \citenamefont {Schindelwig}, \citenamefont {Weidinger},
  \citenamefont {Engleder},\ and\ \citenamefont {F\"urthauer}}]{CodeAcc}%
  \BibitemOpen
  \bibfield  {author} {\bibinfo {author} {\bibfnamefont {Q.}~\bibnamefont
  {Bodini-Lefranc}}, \bibinfo {author} {\bibfnamefont {J.}~\bibnamefont
  {Schindelwig}}, \bibinfo {author} {\bibfnamefont {D.}~\bibnamefont
  {Weidinger}}, \bibinfo {author} {\bibfnamefont {L.}~\bibnamefont
  {Engleder}},\ and\ \bibinfo {author} {\bibfnamefont {S.}~\bibnamefont
  {F\"urthauer}},\ }\href@noop {} {\bibinfo {title} {Arrested coarsening
  code}},\ \bibinfo {howpublished}
  {\url{https://gitlab.tuwien.ac.at/sebastian.fuerthauer/arrestedcoarsening}}
  (\bibinfo {year} {2025})\BibitemShut {NoStop}%
\bibitem [{\citenamefont {Langer}(1971)}]{langer1971theory}%
  \BibitemOpen
  \bibfield  {author} {\bibinfo {author} {\bibfnamefont {J.~S.}\ \bibnamefont
  {Langer}},\ }\bibfield  {title} {\bibinfo {title} {Theory of spinodal
  decomposition in alloys},\ }\href@noop {} {\bibfield  {journal} {\bibinfo
  {journal} {Annals of Physics}\ }\textbf {\bibinfo {volume} {65}},\ \bibinfo
  {pages} {53} (\bibinfo {year} {1971})}\BibitemShut {NoStop}%
\bibitem [{\citenamefont {Novick-Cohen}\ and\ \citenamefont
  {Segel}(1984)}]{novick1984nonlinear}%
  \BibitemOpen
  \bibfield  {author} {\bibinfo {author} {\bibfnamefont {A.}~\bibnamefont
  {Novick-Cohen}}\ and\ \bibinfo {author} {\bibfnamefont {L.~A.}\ \bibnamefont
  {Segel}},\ }\bibfield  {title} {\bibinfo {title} {Nonlinear aspects of the
  cahn-hilliard equation},\ }\href@noop {} {\bibfield  {journal} {\bibinfo
  {journal} {Physica D: Nonlinear Phenomena}\ }\textbf {\bibinfo {volume}
  {10}},\ \bibinfo {pages} {277} (\bibinfo {year} {1984})}\BibitemShut
  {NoStop}%
\bibitem [{\citenamefont {Argentina}\ \emph {et~al.}(2005)\citenamefont
  {Argentina}, \citenamefont {Clerc}, \citenamefont {Rojas},\ and\
  \citenamefont {Tirapegui}}]{argentina2005coarsening}%
  \BibitemOpen
  \bibfield  {author} {\bibinfo {author} {\bibfnamefont {M.}~\bibnamefont
  {Argentina}}, \bibinfo {author} {\bibfnamefont {M.}~\bibnamefont {Clerc}},
  \bibinfo {author} {\bibfnamefont {R.}~\bibnamefont {Rojas}},\ and\ \bibinfo
  {author} {\bibfnamefont {E.}~\bibnamefont {Tirapegui}},\ }\bibfield  {title}
  {\bibinfo {title} {Coarsening dynamics of the one-dimensional cahn-hilliard
  model},\ }\href@noop {} {\bibfield  {journal} {\bibinfo  {journal} {Physical
  Review E—Statistical, Nonlinear, and Soft Matter Physics}\ }\textbf
  {\bibinfo {volume} {71}},\ \bibinfo {pages} {046210} (\bibinfo {year}
  {2005})}\BibitemShut {NoStop}%
\bibitem [{\citenamefont {King}\ and\ \citenamefont
  {Petry}(2020)}]{king2020phase}%
  \BibitemOpen
  \bibfield  {author} {\bibinfo {author} {\bibfnamefont {M.~R.}\ \bibnamefont
  {King}}\ and\ \bibinfo {author} {\bibfnamefont {S.}~\bibnamefont {Petry}},\
  }\bibfield  {title} {\bibinfo {title} {Phase separation of tpx2 enhances and
  spatially coordinates microtubule nucleation},\ }\href@noop {} {\bibfield
  {journal} {\bibinfo  {journal} {Nature communications}\ }\textbf {\bibinfo
  {volume} {11}},\ \bibinfo {pages} {270} (\bibinfo {year} {2020})}\BibitemShut
  {NoStop}%
\bibitem [{\citenamefont {Yan}\ \emph {et~al.}(2022)\citenamefont {Yan},
  \citenamefont {Narayanan}, \citenamefont {Wiegand}, \citenamefont
  {J{\"u}licher},\ and\ \citenamefont {Grill}}]{yan2022condensate}%
  \BibitemOpen
  \bibfield  {author} {\bibinfo {author} {\bibfnamefont {V.~T.}\ \bibnamefont
  {Yan}}, \bibinfo {author} {\bibfnamefont {A.}~\bibnamefont {Narayanan}},
  \bibinfo {author} {\bibfnamefont {T.}~\bibnamefont {Wiegand}}, \bibinfo
  {author} {\bibfnamefont {F.}~\bibnamefont {J{\"u}licher}},\ and\ \bibinfo
  {author} {\bibfnamefont {S.~W.}\ \bibnamefont {Grill}},\ }\bibfield  {title}
  {\bibinfo {title} {A condensate dynamic instability orchestrates actomyosin
  cortex activation},\ }\href@noop {} {\bibfield  {journal} {\bibinfo
  {journal} {Nature}\ }\textbf {\bibinfo {volume} {609}},\ \bibinfo {pages}
  {597} (\bibinfo {year} {2022})}\BibitemShut {NoStop}%
\bibitem [{\citenamefont {Saha}\ \emph {et~al.}(2020)\citenamefont {Saha},
  \citenamefont {Agudo-Canalejo},\ and\ \citenamefont
  {Golestanian}}]{saha2020scalar}%
  \BibitemOpen
  \bibfield  {author} {\bibinfo {author} {\bibfnamefont {S.}~\bibnamefont
  {Saha}}, \bibinfo {author} {\bibfnamefont {J.}~\bibnamefont
  {Agudo-Canalejo}},\ and\ \bibinfo {author} {\bibfnamefont {R.}~\bibnamefont
  {Golestanian}},\ }\bibfield  {title} {\bibinfo {title} {Scalar active
  mixtures: The nonreciprocal cahn-hilliard model},\ }\href@noop {} {\bibfield
  {journal} {\bibinfo  {journal} {Physical Review X}\ }\textbf {\bibinfo
  {volume} {10}},\ \bibinfo {pages} {041009} (\bibinfo {year}
  {2020})}\BibitemShut {NoStop}%
\bibitem [{bow()}]{bowman1953introduction}%
  \BibitemOpen
  \href@noop {} {}\bibinfo {note} {Introduction to Elliptic Functions. By F.
  Bowman. Pp. 115. 12s. 6d. 1953. (English Universities Press)}\BibitemShut
  {NoStop}%
\end{thebibliography}%

\end{document}